\documentclass[prb,aps,preprint]{revtex4}
\usepackage{epsfig}
\usepackage{amsmath}
\usepackage{pst-all}
\usepackage{pstcol}
\draft
\begin{document}
\title{Computing the free energy of molecular solids 
 by the Einstein molecule approach: Ices XIII and XIV, 
hard-dumbbells and a patchy model of proteins}
\author{E. G. Noya, M. M. Conde and 
C. Vega\footnote{published in J. Chem. Phys. 129 104704 (2008)}}
\affiliation{Departamento de Qu\'{\i}mica-F\'{\i}sica, Facultad de
	Ciencias Qu\'{\i}micas, Universidad Complutense de Madrid,
	E-28040 Madrid, Spain}
\date{\today}

\begin{abstract}
	The recently proposed Einstein molecule approach is extended
to compute the free energy of molecular solids. This method is a variant
of the Einstein crystal method of Frenkel and Ladd[J. Chem. Phys. {\bf 81},
3188 (1984)]. In order to show its applicability, we have computed the
free energy of a hard-dumbbells solid, of two recently
discovered solid phases of water, namely, ice XIII and ice XIV, where 
the interactions between water molecules are described by 
the rigid non-polarizable TIP4P/2005 model potential, and 
of several solid phases that are thermodynamically stable for an 
anisotropic patchy model with octahedral symmetry which mimics
proteins.
Our calculations show that
both the Einstein crystal method and the Einstein molecule approach 
yield the same results within statistical uncertainty. 
In addition, we have studied in detail some subtle issues concerning
the calculation of the free energy of molecular solids. 
First, for solids with non-cubic symmetry, we have studied the effect
of the shape of the simulation box on the free energy. Our results show that 
the equilibrium shape of the simulation box must be used to compute
the free energy in order to avoid the appearance of artificial stress 
in the system that
will result in an increase of the free energy.
In complex solids, such as the solid phases of water, another difficulty
is related to the choice of the reference structure. As in some cases
there is not an obvious orientation of the molecules, it is not clear how
to generate the reference structure.  Our results will show that, 
as long as the structure is not too far from the equilibrium structure, 
the calculated free energy
is invariant to the reference structure used in the free energy calculations.
Finally, the strong size dependence of the free energy of solids is also studied.

\end{abstract}


\maketitle
\newpage

\section{Introduction}

Since the pioneering work of Hoover
\emph{et al.},\cite{hoover-ree}
	determining the free energy of molecular solids has been 
an important area of 
research.\cite{frenkel84,mulder,vega98,vlot99,JCP_2004_120_02122,brennecke,sweatman}
One of the most popular methods to compute the free energy of solids
is the Einstein crystal method, proposed by Frenkel and Ladd more than 
two decades ago.\cite{frenkel84}
In this method, the free energy of a given solid is computed by designing
an integration path that links the solid to an ideal Einstein crystal
with the same structure as the real solid, for which the free energy can be 
analytically computed. This method was
soon extended to molecular solids.\cite{mulder} In this case, in addition to the springs
that bound each molecule to its lattice position, springs that keep the
particles in the right orientation must also be added.\cite{mulder,vega_jcp92}
Using this technique, the free energy of several atomic and molecular
solids has been computed.\cite{spherocylinders,spherocylinders2,vega_jcp92,vega_jcp92b,vega98,malanoski,chain,alcanos,bresme00,JCP_2000_112_08534,schroer2000,schroer2001,JCP_2002_117_06313,chain2,JCP_2003_118_00728,vega03,hynninen_pre03,hynninen_jpcm03,tanaka04,PRL_2004_92_255701,sanz2,silica,carbon,fortini06,hynninen,caballero,enrique,noe,sandler}

	Quite recently a new method to compute the free energies of solids
which was denoted as ``the Einstein molecule'' approach 
has been proposed.\cite{vega_noya,review} This method consists of a slight modification 
of the Einstein crystal method.
In the Einstein crystal method, the reference system
is an ideal Einstein crystal with the constraint that the center of mass of
the system is fixed in order to avoid a quasi-divergence in the integral
of the free energy change from the real solid to 
the reference system. This constraint introduces
some complexity in the method. In particular, the derivation of some
terms that contribute to the free energy 
is somewhat involved.\cite{frenkel84,polson}
The main idea behind the Einstein molecule approach is that the
derivation of the analytical expressions can be considerably 
simplified by fixing the position of one molecule instead of 
fixing the center of mass of the system.
The Einstein molecule approach has been successfully applied to compute the free
energy of the hard-spheres (HS) and Lennard-Jones (LJ) face centered cubic (fcc) solids.
Here it will be shown how it can be applied to molecular solids.

	Moreover, even though the Einstein crystal method has been extended
to molecular solids more than twenty years ago, there are several subtle
issues concerning the calculation of the free energy that are not clear yet.
These difficulties are common to the Einstein crystal and Einstein molecule
approaches. One of these issues concerns the shape of the simulation box.
For solids with non-cubic symmetry, prior to the computation of 
the free energy for a given thermodynamic state,
the solid structure must be relaxed to obtain the equilibrium
unit cell corresponding to that thermodynamic state. This is 
not usually a problem in structures with cubic symmetry, as the
equilibrium structure is determined uniquely by the lattice parameter $a$.
However, in structures with lower symmetry, it is convenient to
first perform a simulation in which both the edges and the angles
that define the simulation box are allowed to relax to the equilibrium
structure. 
This can be achieved by, previously to the free energy calculation,
performing a Parrinello-Rahman NpT simulation. Other alternative
would be to perform a simulation at constant volume but where
the shape of the simulation box is allowed to change, i. e.,
a variable-shape constant volume (VSNVT) 
simulation.\cite{spherocylinders2,bolhuis,fortini06,noya08}
We would like to stress the importance of using the equilibrium structure to
compute the free energy, 
otherwise the solid could be under some stress that will
lead to an increase of the free energy. This has already been noted
previously,\cite{mulder,vega_jcp92} but, due to its importance, we 
believe that it is worthy to review this point.

	Another difficulty that one might encounter when computing the free
energy of molecular solids concerns the reference structure that is
used either in the Einstein crystal or in the Einstein molecule approaches.
In simple solids, in which all the particles exhibit the same 
orientation, this does not pose a problem, as the reference 
structure is chosen simply as a solid where all the particles
lie on their lattice positions and are perfectly oriented. However,
for more complex solids, where not all the molecules exhibit
the same orientation, the choice of the reference structure
might be a subtle issue. This is the
case, for example, of some solid phases of water that exhibit 
complex unit cells. In this situation 
several choices are possible. One might choose to build the
reference structure by using experimental
data to obtain the position and orientation of the molecules or,
alternatively, one might choose to perform an energy minimisation,
so that each molecule will be located  
as to minimise the potential energy.\cite{baez_mp95} Another reasonable choice would be
to calculate the average positions and orientations at the particular thermodynamic
state under study. In view
of this ambiguity, it is of interest to investigate the effect
that one choice or another has on the calculation of the 
free energy.

	Finally, another difficulty arises from the strong size
dependence of the free energy of solids. 
In particular, for the fcc HS solid, several authors
have shown that the free energy per particle
decreases linearly with $1/N$, $N$ being the number
of particles in the system.\cite{polson,enrique,vega_noya}
As a consequence, the fluid-solid coexistence point also exhibits a strong size dependence
(note that the finite size effects on the free energy of 
the fluid and on the equation of state of both phases must also be considered).
The size dependence of the fluid-solid coexistence point obtained with
the values of the free energy from those works\cite{vega_noya} is in agreement with the
coexistence points calculated by Wilding and Bruce\cite{wilding00,wilding02} 
using a completely different route, the phase-switch 
Monte Carlo method.\cite{bruce97,wilding_pre00,wilding_rev} 
This strong size dependence has also been observed for other systems, such
as for example, the fcc LJ solid.\cite{barroso,vega_noya}
In this case, the situation is more complicated because,
in addition to the size dependence of the free energy, there is 
also a dependence on the cutoff of the potential. Both effects
must be studied separately.\cite{vega_noya}
This means that in order to perform a rigorous calculation of the free energy
of a given solid, the free energy must be computed for different 
system sizes, so that the value of the free energy at the thermodynamic limit 
can be obtained by extrapolation to $N$ going to infinity.
However, this procedure 
requires performing many simulations to compute
the free energy of a solid at just one thermodynamic state. 
Therefore, it would be useful
to introduce finite size corrections (FSC), i.e.,
a simple recipe that would allow one 
to estimate the value of the free energy in
the thermodynamic limit from simulations of a system
of finite size. In a previous paper, we have proposed 
several FSC whose performances were assessed for 
simple atomic models,
namely, the HS and LJ model potentials.\cite{vega_noya}
The best performance was obtained by the so-called Asymptotic FSC, in which 
the free energy in the thermodynamic limit is estimated from the
free energy at a finite size $N$ by
taking the limit when $N$ tends to infinity
in the expression used to compute the free energy. 
Depending on how this limit was taken, three different variants
were proposed, and all of them give quite reasonable estimates
of the free energy in the thermodynamic limit.
However, these results might not be general and, therefore,
it would be of interest to check whether the FSC work well also for
molecular solids.

	In this paper we will address all these issues concerning the computation
of free energy of solids. It is our hope that this will contribute to 
encourage other authors to compute free energies.
The paper will be structured in the following way. First, the recently
proposed Einstein molecule approach will be extended to the case of 
molecular solids and it will be shown that the results obtained for all
the solids studied (i.e., a hard-dumbbells solid, a solid
made of anisotropic particles with octahedral symmetry
and the two recently discovered solid phases of water, ice XIII and
ice XIV) are in agreement,
within statistical uncertainty, with the results obtained with the
Einstein crystal method. 
Second, the free energy of ices XIII and XIV
using the rigid non-polarizable TIP4P/2005
model of water will also be calculated. These calculations will serve
to illustrate the importance of obtaining the equilibrium shape of the 
simulation box previously to the computation of the free energy and to
explore what is the best choice for the reference structure that is used
in the computation of the free energy.
Finally, we will perform a systematic study of the size dependence of the free energy
of several crystalline solids for a simple anisotropic patchy model
with octahedral symmetry. The performance of the previously proposed FSC
will be assessed for this model. 

%
\section{Method}
\label{method}
%

\subsection{Model potentials and solid structures}

	In what follows we will consider several pair potentials,
for which the intermolecular potential will be expressed as:
\begin{equation}
U_{sol}=\sum_{i=1}^{N-1}\sum_{j=i+1}^N u_{sol} (i,j)
\label{eq_usol}
\end{equation}
where $u_{sol}(i,j)$ is the intermolecular potential
between molecules $i$ and $j$.

\subsubsection{Hard-dumbbells}

	The first model we considered is the hard-dumbbells (HD) model,
in which each particle consists of two hard-spheres, each of diameter $\sigma_{HS}$,
separated by a distance $L$.
The free energy of this model has
already been studied previously using the Einstein crystal method\cite{vega_jcp92,dijkstra08}
and also theoretically using an extension of the Wertheim theory.\cite{macdowell,vega_jml} 
The possible solid structures for hard-dumbbells have already been 
been discussed in previous works.\cite{vega_jcp92,vega97,dijkstra08}
Hard-dumbbells can form a hexagonal lattice by arranging the dumbbells
in such a way that each
sphere of a dumbbell lies in a hexagonal layer. The dumbbell axis is then tilted from
the normal to the layer by an angle equal to $arcsin(\frac{L}{\sigma_{HS}\sqrt{3}})$.
These layers can be stacked
as to form a fcc lattice (structure designated as CP1) or a hcp
lattice (structure CP2). In these two structures all the dumbbells
exhibit the same tilt angle. Another structure can be obtained by
stacking the layers in such a way that the tilt angle alternates 
between adjacent layers (structure designated as CP3). It has been
shown that only the CP1 structure is thermodynamically stable 
(for $L^* = L/\sigma_{HS} > 0.4$).\cite{dijkstra08}
For values of $L^*$ lower than approximately 0.4,
there is a range of pressures for which a plastic fcc crystal is the most stable
phase.\cite{vega97,dijkstra08}
Finally, it is also possible to form aperiodic fcc and hcp structures, i.e., structures
in which the axis of the hard-dumbells are not 
aligned.\cite{wojciechowski,vega_jcp92b,vega_jml} The aperiodic fcc structure becomes
thermodynamically stable for values of the elongation $L^*$
close to unity.\cite{wojciechowski,vega_jcp92,vega_jml,dijkstra08}  
As the main purpose is to show that both the Einstein molecule approach and
the Einstein crystal method lead to the same value of the free energy, we have
chosen to study only the structure designated as CP1
for hard-dumbbells with $L^*=1$ (for
this elongation the CP1 solid is metastable).

\subsubsection{The TIP4P/2005 water model. Ices XIII and XIV}

	The interaction between water molecules was modelled using
a rigid non-polarizable model potential, the TIP4P/2005 water model.\cite{abascal05b}
This model is a variant of the TIP4P potential,\cite{jorgensen83}
in which the water molecule is modelled by one LJ
interaction site on the oxygen atom, two positive charges located on
the hydrogen atom and a negative charge that is located on the H-O-H 
bisector. It has been shown that the TIP4P model is able to predict reasonably well the phase
diagram of water. It predicts that ice Ih is the most stable solid
phase at the normal melting point and it reproduces the densities
of the solid phases of water within 2\% of the experimental values.\cite{pccpgdr}
The main failure of this method seems to be a melting point about 40K
below the experimental value.\cite{PRL_2004_92_255701,ramon06} It was then clear that
the model could be improved
and several groups proposed variants of
this model. In particular, the TIP4P/Ice model\cite{abascal05a} 
has been fitted to reproduce the experimental melting
point of water and the TIP4P/Ew\cite{JCP_2004_120_09665} and 
TIP4P/2005\cite{abascal05b} models reproduce 
the maximum in density at room pressure.
Among these models, we have chosen to use the TIP4P/2005 model potential,
because it  provides a good description of the 
phase diagram\cite{abascal05b} and also it predicts 
to good accuracy the density of the solid phases of ice.\cite{eos}

	In this work, the free energies of two recently 
discovered solid phases of water, namely ices XIII and XIV,\cite{iceXIII} 
are computed for the first time.
Ice XIII is the proton ordered form of ice V. It has a monoclinic
unit cell with 28 molecules. Ice XIV is the
proton ordered form of ice XII. It has a tetragonal unit cell
with 12 molecules. 
The TIP4P/2005 model has been shown to reproduce reasonably well
the densities of these two solid forms of ice.\cite{maria06b}
In the simulations performed in this work the LJ potential was truncated at 8.5 {\AA}
for both solid phases.
Standard long range corrections were added to the LJ energy.\cite{allen_book,frenkelbook}
Ewald sums were used to deal with the long range electrostatic forces.
The real part of the electrostatic contribution was also truncated at 8.5 \AA~.
The screening parameter and the number of vectors of reciprocal
space considered had to be carefully selected for each
crystal phase.\cite{allen_book,frenkelbook}

\subsubsection{Model particles with octahedral symmetry}

	We also computed the free energy
of a patchy model, which has been previously
used as a simplified model of globular proteins.\cite{jon,alex,eva_patchy}
This model consists of a repulsive core with some attractive sites (patches) on
its surface. In particular, we studied model particles with six patches
in an octahedral arrangement. 
The repulsive core is modelled by the LJ repulsive 
core, while the attractive term is described by the LJ tail modulated
by Gaussian functions centred at the positions of each patch. Therefore,
the total energy between two particles is described by the following
function:
\begin{equation}
u_{patchy}({\bf r}_{ij},{\bf \Omega}_i,{\bf \Omega}_j)  = 
\begin{cases}
u_{LJ}(r_{ij})  & r_{ij} <  \sigma_{LJ} \\
u_{LJ}(r_{ij}) \exp \left(-\frac{\theta_{k_{min},ij}^2}{2\sigma^2 } \right)
         \exp \left(-\frac{\theta_{l_{min},ji}^2}{2\sigma^2 } \right)
& r_{ij} \ge  \sigma_{LJ}
\end{cases}\label{eq_patchy}
\end{equation}
where $u_{LJ}(r_{ij})$ is the Lennard-Jones potential,
$\sigma$ is the standard deviation of the Gaussian,
$\theta_{k,ij}$ ($\theta_{l,ji}$) is
the angle formed between patch $k$ ($l$) on atom $i$ ($j$) and
the interparticle vector
${\bf r}_{ij}$ ( ${\bf r}_{ji}$), and $k_{min}$ ($l_{min}$) is the patch
that minimises the magnitude of this angle.
Additionally, for computational efficiency, the potential is truncated and
shifted using a cutoff distance of $2.5\,\sigma_{LJ}$.

	Using reduced units (i.e., choosing the unit of energy and length
as the values of the LJ parameters $\varepsilon_{LJ}$ and $\sigma_{LJ}$), the only 
parameter that needs to be specified is the width of the patches $\sigma $.
In this work, we have chosen $\sigma =0.3$ rad., as for this value 
the whole phase diagram has already been studied.\cite{eva_patchy}
In this previous study, it has been shown that there are several solid
phases that are thermodynamically stable, namely, simple cubic (sc), 
body-centred cubic (bcc), face-centred
cubic (fcc) and, at high temperatures, a plastic fcc crystal.
In this work, we will compute the free energy of the three orientationally
ordered structures (sc, bcc, and fcc) for several system sizes.

\subsection{The Einstein molecule approach for molecular solids}

	As mentioned in the Introduction, the Einstein molecule approach 
is a variant of the Einstein crystal
method of Frenkel-Ladd\cite{frenkel84} that has been proposed quite recently.\cite{vega_noya}
Analogously to the Frenkel-Ladd method, the free energy is computed by
integration to a reference system whose free energy 
can be computed analytically. The difference is that in the Einstein
molecule approach the reference
system is not an ideal Einstein crystal, but an ideal Einstein molecule. The 
Einstein molecule is defined as an Einstein crystal in which
one of the particles does not vibrate. The name of Einstein molecule has been
chosen by analogy with molecules, where it is common to use one
of the atoms to define the position of a molecule, and the vibrational movement
of the remaining atoms is given relative to this reference atom.
The Einstein molecule approach has been successfully applied to compute the free energy
of simple atomic systems 
(HS and LJ),\cite{vega_noya} 
but we will see that it can be easily extended
to molecular solids.

	We will start by writing the partition function 
of a molecular system in the canonical ensemble:
\begin{equation}
Q= \frac{ q'^{N} }{N! \Lambda^{3N}}
\int \exp\left[-\beta U({\bf r}_1,\omega_1,...,{\bf r}_N,\omega_N) 
\right]d{\bf r}_1 d\omega_1...d{\bf r}_N d\omega_N
\label{eq_pf}
\end{equation}
where ${\bf r}_i=(x_i, y_i, z_i)$ is the position of the reference point of molecule
$i$ in Cartesian coordinates, and $\omega_i$ stands for a set of normalised
angles (i.e., $\int d\omega_i=1$) defining the orientation of particle $i$.
$q'=q_rq_vq_e$, where $q_r$, $q_v$ and $q_e$ are the rotational, vibrational,
and electronic partition functions, respectively. 
$\Lambda $ is the thermal de Broglie wavelength ($\Lambda=(h^2/(2\pi mk_BT))^{1/2}$).
There is some
freedom in choosing the reference point of the molecule. It
can be chosen as the center of mass or, alternatively, this reference point
can be chosen so that all elements of symmetry pass through it
(for a more detailed discussion see Ref. \onlinecite{review}).
We have chosen the reference point to be at the center of the sphere 
for the octahedral patchy model and for spheric particles, 
at the center of mass for hard-dumbbells and at the oxygen
atom for water.

The intermolecular potential $U$ depends only on the relative distance between the molecules,
not on their absolute positions, i.e., it is invariant under translations. This invariance
of the system can be used to write the partition function in a more convenient way by
performing a change of variables from $({\bf r}_1,{\bf r}_2, ...,{\bf r}_N)$
to $({\bf r}_1, {{\bf r}_2}^{'}={\bf r}_2-{\bf r}_1, ...,  {{\bf r}_N}^{'}={\bf r}_N-{\bf r}_1)$.
Therefore, Equation \ref{eq_pf} can be written:
\begin{eqnarray} 
Q & = & \frac{ q'^{N} }{N! \Lambda^{3N}} \int d{\bf r}_1
\int \exp\left[-\beta  U(\omega_1,{\bf r}'_2,\omega_2,...,{\bf r}'_N,\omega_N) \right] 
d\omega_1 d{\bf r}'_2 d\omega_2...d{\bf r}'_N d\omega_N  \nonumber \\
& = & \frac{q'^{N}}{N! \Lambda^{3N}}
\int d{\bf r}_1 \, \kappa
\label{eq_inte}
\end{eqnarray}
The integral $\kappa$ does not depend on the position of particle 1, ${\bf r}_1$.
Therefore, 
the integration over ${\bf r}_1$ can readily be performed:
\begin{equation}
Q=\frac{q'^{N}}{N! \Lambda^{3N}} V \, \kappa
\end{equation} 
For a system of $N$ indistinguishable particles and for
a given position of particle 1 there are $(N-1)!$ possible 
permutations of the remaining
$N-1$ particles. The term $\kappa$ can be evaluated by computing the
integral for a given permutation of the particles ($\kappa'$) and multiplying it by the number of 
permutations, so that the partition function can be written as:
\begin{equation}
Q= \frac{q'^N}{N! \Lambda^{3N}} V (N-1)! \hspace{0.2cm} \kappa'
= \frac{q'^N}{N\Lambda^{3N}} V \, \kappa'
\label{eq_general_partition}
\end{equation}
We will assume that $q'$
has the same value in the two coexisting phases, so 
that its value does not affect the coexistence point. For simplicity,
in what follows, we will assign $q'$ the value unity.

We will extend now the definition of the ideal Einstein molecule to molecular solids.
For atomic solids, an ideal Einstein molecule was defined as an ideal
Einstein crystal in which one of the particles does not vibrate
and acts as reference.
For molecular solids, the ideal Einstein 
molecule is defined as an ideal Einstein crystal in which
the reference point of particle 1 is fixed, but rotations of
the molecule about this point are allowed.
The reference point of particle 1 is called the carrier, because it transports the lattice, i.e.,
the position of the lattice is uniquely defined by the position of 
the reference point of particle 1.
The lattice can move as a whole over the volume of the simulation box, 
and its position is defined by the position of the reference point of
particle 1. The potential energy of the
ideal Einstein molecule is given by:
\begin{eqnarray} 
\label{ete_einstein_molecule}
U_{Ein-mol-id} & = & U_{Ein-mol-id,t}+U_{Ein,or}       \nonumber \\
U_{Ein-mol-id,t} & = & \sum ^{N}_{i=2} u_{Ein-mol-id,t}
= \sum ^{N}_{i=2} \left[\Lambda_E({\bf r}_{i}-{\bf r}_{io})^{2} \right] \\
U_{Ein,or} & = & \sum ^{N}_{i=1} u_{Ein,or} \nonumber
\end{eqnarray}
where ${\bf r}_{io}$ is the position of the reference point of molecule $i$
in the reference Einstein solid, while ${\bf r}_i$ represents its position in
the current configuration.
As can be seen in Eq.\ref{ete_einstein_molecule}, 
all the particles except particle 1 (which is fixed) are attached to 
their lattice positions by harmonic springs. 
An orientational field ($U_{Ein,or}$) that forces the particles to adopt the right
orientation is also included
(this field acts over all the particles of the system, including particle 1).
The orientational field depends on the symmetry of the particles and, thus,
an orientational field must be defined for each model potential. 
The orientational field used for each one of the model potentials that
have been studied in this work will be given in Section \ref{orient}.

	The partition function of the ideal Einstein molecule can be obtained
by performing the integral $\kappa'$ for this particular case:
\begin{eqnarray}
\kappa'_{Ein-mol-id} & = & \left[ \int \exp{[-\beta \Lambda_E 
({\bf r}-{\bf r}_{0}})^2] d{\bf r} \right]^{(N-1)} 
\left[ \int \exp{(-\beta u_{Ein,or})} d\omega \right]^N = \nonumber \\
& = & \left( \frac{\pi}{\beta \Lambda_E} \right)^{3(N-1)/2}  Q_{Ein,or}
\label{kprima}
\end{eqnarray}
 where $Q_{Ein,or}$ is the orientational partition function,
which is usually evaluated numerically (more details are given
the Section \ref{orient}). 

The free energy of the ideal Einstein molecule
can be obtained by replacing the partition function given by Eq. \ref{kprima} 
in Eq. \ref{eq_general_partition}:
\begin{eqnarray} 
\frac{\beta A_{Ein-mol-id}}{N} & = & \frac{\beta A_{0}}{N} = 
\frac{\beta A_{0,t}}{N} + \frac{\beta A_{0,or}}{N} = - \frac{1}{N} ln (Q) \nonumber \\
 & = &  
\left[ \frac{1}{N} \, ln \left( \frac{N\Lambda^{3}}{V} \right) + \frac{3}{2}
\left(1-\frac{1}{N}\right) \hspace{0.2cm} ln \left( 
\frac{ \Lambda^ 2 \beta \Lambda_E }{\pi} \right) \right]
+ \left[ -\frac{1}{N} ln ( Q_{Ein,or} ) \right]
\label{eq_free_energy}
\end{eqnarray}
The numeric value of the thermal de Broglie wavelength $\Lambda$ is irrelevant
to compute the coexistence point as long as the same value is used for both
coexisting phases. Therefore, we have chosen to assign $\Lambda$ the value
of the characteristic length for each model potential.
Thus, for HS $\Lambda=\sigma_{HS}$, for HD  $\Lambda=\sigma_{HS}$,
for LJ $\Lambda=\sigma_{LJ}$, for water $\Lambda=1$ {\AA} and for
the patchy model $\Lambda = \sigma_{LJ}$.

	In the Einstein molecule approach, the free energy of a given solid
is estimated by designing a path from the ideal Einstein molecule (whose free energy
can be computed by Eq. \ref{eq_free_energy}) to the real solid. This path can be
divided into three steps (see Figure \ref{fig_esquema}). In the first step, the ideal Einstein molecule
or, what is the same, the position of the reference point of the carrier (molecule 1) is
constrained to a given position. 
In the second step,
the ideal Einstein molecule with fixed molecule 1 is transformed into the real
solid with fixed molecule 1. Finally, in the last step, the solid of interest is recovered
by removing the constraint over the position of molecule 1. 
The free energy change that results from the transformation in the first step
is given by a term $k_BTln(V/\Lambda^3)$, while the third step 
contributes by a term $-k_BTln(V/\Lambda^3)$. 
The term
$V$ comes from the constraint on the position of molecule 1, and the
term $\Lambda^3$ comes from the constraint on the momentum.
Therefore, the contributions to the final free energy of steps one and three cancel out and
all what is needed is to compute the free energy change between an ideal Einstein molecule
and the real solid, both with the position (but not the orientation)
of particle 1 fixed. This free energy change
will be computed in two stages. In the first stage we will evaluate the free energy change
between the ideal Einstein molecule (there is no interaction between the particles,
only the external Einstein crystal field is present) and the interacting Einstein 
molecule (in which both the
springs and the intermolecular potential are present), both with the position of particle
1 fixed, by a perturbative approach:\cite{zwanzig}
\begin{equation} 
\label{eq_deltaa1} 
\Delta A_{1}=  U_{lattice}-k_BT \ln
\left<\exp\left[-\beta(U_{sol}-U_{lattice})\right]\right>_{Ein-mol-id}.
\end{equation}
where $U_{sol}$ is the potential energy of the real solid and
$U_{lattice}$ is the potential energy of the frozen lattice
(see Ref. \onlinecite{review} for a more detailed discussion).
The brackets with the subscript $Ein-mol-id$ indicate that the average is performed
by sampling the configurations in a system where only the Einstein field is
present.
In the second stage, the interacting Einstein molecule with fixed molecule
1 is transformed into the real
solid with fixed molecule 1, by slowly turning off the springs, 
according to the following expression:
\begin{equation}
U(\lambda)=\lambda U_{sol}+( 1 - \lambda) ( U_{Ein-mol-id} + U_{sol} )
\end{equation}
where $\lambda$ is a parameter that takes values between 0 and 1.
The free energy change corresponding to this transformation can be estimated
by numerically evaluating the following integral:
\begin{equation} 
\label{eq_deltaa2}
\Delta A_2 =-\int_0^{\Lambda_E} \frac{\left<U_{Ein-mol-id}\right>_{N,V,T,\lambda}}{\Lambda_E} 
d(\lambda \Lambda_E ).
\end{equation}
This integral is usually performed by using a Gauss-Legendre quadrature formula.
For that purpose, the integrand of this expression must be evaluated
at several values of $\lambda \Lambda_E$, which can be done by performing
$NVT$ MC simulations for those values of the coupling parameter.

Taking all the contributions together, the free energy of solid can be computed as:
\begin{equation}        
\label{a_total}
A_{sol}=A_{Ein-mol-id} + \Delta A_1 + \Delta A_2 
\end{equation}
	which is the central result of this work.

	An alternative proof of the Einstein molecule approach can be found 
in the Appendix. We show that the Einstein molecule method can be obtained
as the limit case of the Einstein crystal method when the mass of molecule 1
is much larger than the mass of the remaining molecules.


\subsection{Free energy of the orientational field}
\label{orient}

	We have said before that the orientational field
must be chosen so that it has the same symmetry as the molecules. 
In this section, the orientational fields used for each of
the studied model potentials are given.
In particular, for hard-dumbbells ($D_{\infty ,h}$ symmetry), we
have chosen the orientational field:\cite{vega_jcp92}
\begin{equation}
U_{Ein,or}=\sum ^{N}_{i=1} \left[ \Lambda_{E,b}\sin^2\left( \psi_{b,i} \right)
\right].
\end{equation}
where $\psi_{b,i}$ is the angle formed between the axis of particle $i$ and 
the equilibrium position of the axis of particle $i$ in the  
CP1 HD solid. 
In this case, the partition function of the orientational field 
can be computed as:
\begin{equation}
Q_{Ein,or}=\left[ \frac{1}{4\pi}
\int exp \left( - \beta \Lambda_{E,b} sin^2(\psi_{b,i}) 
\right) sin \theta d\theta d\phi \right]^N
\end{equation}
	where $\theta$ and $\phi$ are the polar angles that define the
orientation of the axis of the molecule.
In this case, the angle $\psi_{b,i}$ can be identified with the polar angle $\theta$.
Therefore, this expression can be simplified to the following integral
in one dimension:
\begin{equation}
Q_{Ein,or}=\left[ \int_0^1 exp[\beta\Lambda_{E,b} (x^2-1)] dx \right]^N
\end{equation}
This integral can be evaluated using a numerical integration
method, such as, for example, the Simpson's rule.

The water molecule exhibits $C_{2v}$ symmetry and, therefore, a convenient
choice of the orientational field is:\cite{vega98}
\begin{equation}
U_{Ein,or} =   \sum^{N}_{i=1} \left[ \Lambda_{E,a} \sin^2 \left(\psi_{a,i} \right) +
\Lambda_{E,b} \left( \frac{\psi_{b,i}}{\pi} \right)^2 \right].
\end{equation}
In this case, the orientation of the molecule is defined by two unitary
vectors, $\vec{a}$ and $\vec{b}$. These vectors are obtained as the
subtraction ($\vec{a}$) and the addition ($\vec{b}$) of the two hydrogen 
vectors given relative to the position
of the oxygen atom. The angle $\psi_{a,i}$ is the angle
formed by the vector $\vec{a}$ of molecule $i$ in a given configuration ($\vec{a}_i$) 
and the vector $\vec{a}$ in the reference configuration ($\vec{a}_{i,0}$)
of the external Einstein field.
$\psi_{b,i}$ is defined analogously but with vector $\vec{b}$ (for further
details see Ref. \onlinecite{review}).
The orientational partition function can be computed as:
\begin{equation}
Q_{Ein,or}= \left[ \frac{1}{8\pi^2} \int exp \left( -\beta
\left\{ \Lambda_{E,a}sin^2(\psi_a)+\Lambda_{E,b} \left( \frac{\psi_b}{\pi}\right)^2 
\right\} \right)
sin\theta d\theta d\phi d\chi \right]^N
\end{equation}
	where $\theta$, $\phi$ and $\chi$ are the Euler angles that
define the orientation of the molecule.
This integral can be simplified by choosing the vector $\vec{b}_{0}$ as the $z$ axis,
so that the Euler angle $\theta $ is identical to the vector $\psi_b$. 
It can be evaluated numerically by using a Monte Carlo integration method.
The efficiency of the Monte Carlo integration method can be considerably 
improved by realizing that, for large values of $\Lambda_{E,b}$,
the exponential decays very rapidly to zero as the angle $\theta $ increases,
i.e., as the particle rotates away from the reference orientation.
Therefore, much efficiency is gained by sampling only small values of $\theta $.
We have chosen to sample cos$\theta$ and only those
angles for which the cosine is between 0.99 and 1 have been considered.
About $5000\times10^6$ MC cycles were used to evaluate this integral.
In a previous paper, it has been shown that some approximations can be
made to this integral for large values of the coupling parameter.\cite{vega98} 
We have found that, for a coupling parameter 
$\Lambda_{E,a}/(k_BT)=\Lambda_{E,b}/(k_BT)=$25000, the approximation gives
a value for the free energy of the orientational field,
$A_{Ein,or}/(Nk_BT)=-1/Nln(Q_{Ein,or})$, about 0.04 lower than that obtained
by performing the exact integral using the Monte Carlo integral method. 
In particular, using the exact integral we obtained that 
$A_{Ein,or}/(Nk_BT)=16.05$, while using the approximate formula,
we obtained that $A_{Ein,or}/(Nk_BT)=16.01$.
Although this difference is not too large,
we recommend to use the
exact expression of the integral, using
a numerical algorithm to evaluate it.

As with regard to the patchy model with octahedral symmetry (point group
$O_h$), the 
orientational field was:\cite{eva_patchy}
\begin{equation}
U_{Ein,or}=\sum ^{N}_{i=1} \left[
\Lambda_{E,a}\sin^2\left(\psi_{a,i,min}\right)
+ \Lambda_{E,b}\sin^2\left( \psi_{b,i,min} \right)
\right].
\end{equation}
where $\psi _{a,i,min}$ is the minimum angle formed by any of  
the vectors that define the position of the patches
in the particle's reference system
with respect to the $x$ axis of a fixed reference system
and $\psi_{b,i,min}$ is the analogous quantity with 
respect to the $y$ axis, where the fixed reference system
has been chosen to be
coincident with the orientation of the patches in the
perfect lattice. 
Therefore, the orientational partition function 
is given by:
\begin{equation}
Q_{Ein,or}=\left[ \frac{1}{8\pi^2} \int exp \left\{ -\beta (\Lambda_{E,a} 
sin^2(\psi_{a,min}) + \Lambda_{E,b} sin^2 (\psi_{b,min}) ) \right\}
sin\theta d\theta d\phi d\chi \right ]^N
\end{equation}
In this case, the integral was evaluated numerically using the Monte
Carlo integration method and using at least $10^9$ points.

In all the cases, we have chosen that both the translational and orientational field have 
the same numeric value of the coupling parameter 
$\Lambda_E=\Lambda_{E,a}=\Lambda_{E,b}$. Note, however, that the coupling 
parameter of the translational field, $\Lambda_E$ has units of energy
over a squared length, whereas
the orientational coupling parameters $\Lambda_{E,a}$ and $\Lambda_{E,b}$
have dimensions of energy.

	Once the orientational field has been chosen, we can write
the explicit form for the integral $\Delta A_2$. For example, for 
water:
\begin{equation} 
\label{eq_deltaa2_water}
\Delta A_2 =-\int_0^{\Lambda_E} \left< 
\sum ^{N}_{i=2} \left({\bf r}_{i}-{\bf r}_{io}\right)^{2} +
\sum^{N}_{i=1} \left[ \sin^2 \left(\psi_{a,i} \right) +
\left( \frac{\psi_{b,i}}{\pi} \right)^2 \right]
\right>_{N,V,T,\Lambda^{'}}
d\Lambda'
\end{equation}
where the brackets with the subscript $N,V,T,\Lambda^{'}$ means an
average over a simulation of a system where both an ideal Einstein field with coupling
parameter $\Lambda^{'}$ (where $\Lambda^{'}=\lambda \Lambda_E$)
and the solid potential are present
(i.e., the total potential is 
$U_{sol}+\Lambda^{'}\sum ^{N}_{i=2} \left({\bf r}_{i}-{\bf r}_{io}\right)^{2} +
\Lambda^{'} \sum^{N}_{i=1} ( \sin^2 (\psi_{a,i}) +
( \frac{\psi_{b,i}}{\pi} )^2 )$). For convenience, we will split this
expression in two terms, one that accounts for the 
translational contribution ($\Delta A_{2,t}$)
and other that accounts for the orientational 
contribution ($\Delta A_{2,or}$):
\begin{equation} 
\Delta A_{2,t} =-\int_0^{\Lambda_E} \left< 
\sum ^{N}_{i=2} \left({\bf r}_{i}-{\bf r}_{io}\right)^{2}
\right>_{N,V,T,\Lambda^{'}}
d\Lambda'
\label{eq_a2t}
\end{equation}
\begin{equation} 
\Delta A_{2,or} =-\int_0^{\Lambda_E} \left< 
\sum^{N}_{i=1} \left[ \sin^2 \left(\psi_{a,i} \right) +
\left( \frac{\psi_{b,i}}{\pi} \right)^2 \right]
\right>_{N,V,T,\Lambda^{'}}
d\Lambda'
\label{eq_a2or}
\end{equation}

\subsection{Finite size corrections}

	It is well known that the free energy of solids exhibits a strong size
dependence.\cite{frenkel84,polson,barroso,vega_noya,enrique} 
In a recent paper, we have made an attempt to propose some recipes 
to correct for this strong size dependence in a simple way. 
In what follows, we briefly review those FSC (a more detailed
discussion was already given in Ref. ~\onlinecite{vega_noya}).

	The two first proposed FSC, namely, the Frenkel-Ladd FSC (FSC-FL) and
the hard-spheres FSC (FSC-HS) consist of simply adding a term
to the free energy of a system of $N$ particles to obtain an approximation
to the free energy in the thermodynamic limit:
\begin{equation}
   \frac{A_{FSC-FL}}{Nk_BT} (N \to \infty)  \simeq  \frac{A_{solid}(N)}{Nk_BT} + \frac{2 ln N}{N}
\end{equation}
\begin{equation}
   \frac{A_{FSC-HS}}{Nk_BT} (N\to \infty)  \simeq  \frac{A_{solid}(N)}{Nk_BT} + \frac{7}{N}
\end{equation}
	These are empiric corrections that have been shown to improve the results
for the HS fcc solid. Also we have noted that the term $\frac{3}{2N}lnN$ is approximately
equal to the term $7/N$ except for very small values of $N$. Therefore, we decided
to explore also the performance of this FSC:
\begin{equation}
   \frac{A_{FSC-HS2}}{Nk_BT} (N\to \infty)  \simeq  \frac{A_{solid}(N)}{Nk_BT} + \frac{3}{2}
\frac{lnN}{N}
\end{equation}

        In a second family of FSC which was designated as FSC-Asymptotic,
 the free energy in the thermodynamic limit
is estimated by taking the limit when $N$ tends to infinity
in the analytical expression of the free energy of the ideal Einstein molecule
(Eq. \ref{eq_free_energy}).
Three different variants of the FSC-Asymptotic
were proposed differing on whether a further approximation to the term
$\Delta A_2$ was also made. In the first variant ($A_{FSC-as1}$), no approximation
was made to $\Delta A_2$:
\begin{equation}
   \frac{A_{FSC-as1}}{Nk_BT} (N\to \infty) \simeq  \frac{3}{2} ln
\left( \frac{ \Lambda^2 \beta \Lambda_E }{\pi} \right) +
\frac{A_{0,or}}{Nk_BT} +
\frac{\Delta A_1(N,\Lambda_E)}{Nk_BT} +
\frac{\Delta A_2(N,\Lambda_E)}{Nk_BT} 
\label{eq_bsc_a1}
\end{equation}
        In a second variant, an approximation to $\Delta A_2$ is made
based on the assumption that all the $N-1$ oscillators contribute by
the same amount to the integral. This is a reasonable approximation for atomic
solids (for a fcc HS solid with N=108 particles, we obtained that all the atoms except the
first nearest neighbours contributed approximately by the same amount;
the contribution of the nearest neighbours is about a 10\% lower than
the contribution of the remaining atoms).
For molecular solids, it is important to
notice that there are two contributions to $\Delta A_2$, one
translational and one orientational. As pointed out before, the
Einstein molecule only imposes the constraint on the position of
particle 1, but not on its orientation. Therefore, assuming that
all the molecules contribute by the same amount to the translational
integral, $\Delta A_2$
can be approximated by the following expression:
\begin{equation}
\frac{\Delta A_2}{Nk_BT} = \frac{\Delta A_{2,t}}{Nk_BT} + \frac{\Delta A_{2,or}}{Nk_BT} =
\frac{N-1}{N} I_{t} + \frac{\Delta A_{2,or}}{Nk_BT} 
= \left( 1-\frac{1}{N} \right) I_{t} + \frac{\Delta A_{2,or}}{Nk_BT}
\end{equation}
       where $\Delta A_{2,t}$ and $\Delta A_{2,or}$ are given
in Eq. \ref{eq_a2t} and \ref{eq_a2or}, and
$I_t$ is the contribution to the translational 
integral of one single arbitrary particle (under the assumption 
that all the particles contribute by the same amount). We shall
assume now that the 
orientational contribution is independent of the system
size and that the asymptotic value of $\Delta A_{2,t}/Nk_BT$ is
$I_t$. In the FSC-as2, $\Delta A_2$ is approximated as:
\begin{equation}
\frac{\Delta A_2}{Nk_BT} (N\to \infty) \simeq I_t + \frac{\Delta A_{2,or}}{Nk_BT}
\end{equation}
Therefore, the FSC-as2 for molecular solids must be slightly modified
with respect to that obtained for atomic solids (compare with Eq. 35 of
Ref. \onlinecite{vega_noya}):
\begin{eqnarray}
  \frac{A_{FSC-as2}}{Nk_BT} (N\to \infty) & \simeq & \frac{3}{2} ln \left( 
\frac{ \Lambda^2  \beta \Lambda_E }{\pi} \right)
+ \frac{A_{0,or}}{Nk_BT} 
+  \frac{\Delta A_1(N,\Lambda_E)}{Nk_BT} \nonumber \\
& + & \frac{\Delta A_{2,or}(N,\Lambda_E)}{Nk_BT} + 
\frac{\Delta A_{2,t} (N,\Lambda_E)/(Nk_BT)}{(1-1/N)}
\label{eq_bsc_a2}
\end{eqnarray}
 	Finally, the last variant is obtained as the mean value of the FSC-as1 and FSC-as2:
\begin{eqnarray}
  \frac{A_{FSC-as3}}{Nk_BT} (N\to \infty) & \simeq &   
\frac{3}{2} ln \left( \frac{ \Lambda^2 \beta \Lambda_E }{\pi} \right) +
 \frac{A_{0,or}}{Nk_BT} +
\frac{\Delta A_1 (N,\Lambda_E)}{Nk_BT} + \frac{\Delta A_{2,or}(N,\Lambda_E)}{Nk_BT} + \nonumber \\
& & \frac{1}{2} \left( \frac{\Delta A_{2,t}(N,\Lambda_E )}{Nk_BT} +  
\frac{\Delta A_{2,t}(N,\Lambda_E )/(Nk_BT)}{(1-1/N)} \right)\label{eq_bsc_a3}
\end{eqnarray}
Notice that in these expressions $\Delta A_1$ and $\Delta A_2$ were obtained
by the Einstein molecule approach.

\section{Results}
\label{results}
%

\subsection{The Einstein molecule approach}

\begin{figure}[!ht]
\begin{center}
\includegraphics[width=85mm,angle=0]{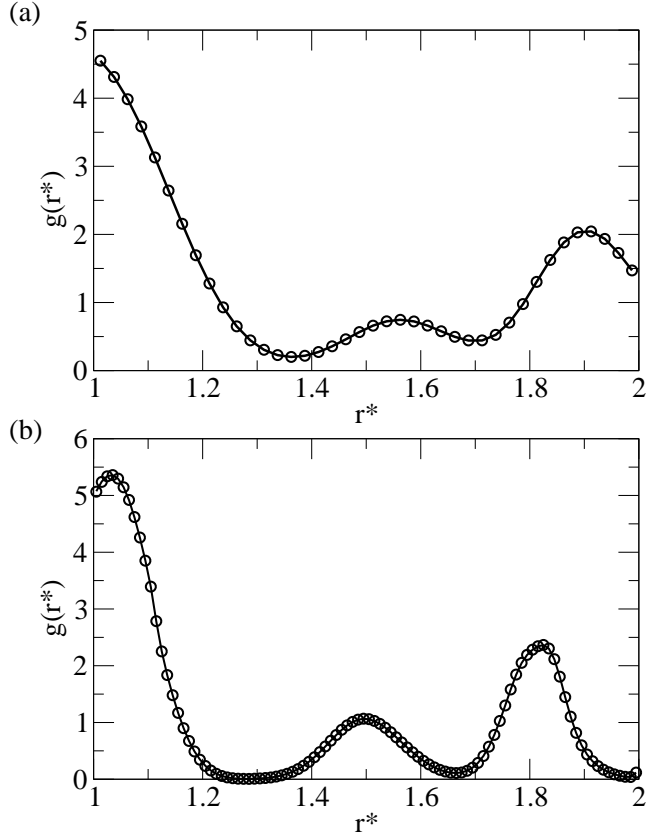}
\caption{\label{fig_fdr} (a) Radial distribution function for hard spheres in the fcc solid
phases when particle 1 moves (solid line) and when it does not move (open circles).
Results correspond to $\rho^{*}=1.04086$ for a system size $N=108$. (b) Site-site 
radial distribution function for hard-dumbbells
with $L^*=1$ in the CP1 structure at $\rho^{*}=0.590$ and for a system size $N=32$
when molecule 1 moves (solid line)
and when the 
reference point of molecule 1 (i.e., its center of mass) is fixed
but molecule 1 can rotate (open circles). As can be seen, the structural
properties are the same, illustrating that the properties of 
the solid present translational invariance.}
\end{center}
\end{figure}

	Before presenting the results of the free energy calculations 
with the Einstein molecule approach, we will show that fixing
one molecule in a solid (in the absence of the Einstein field)
does not affect the structural properties
(due to the translational invariance).
For that purpose, we computed the radial distribution
function in a NVT simulation for an atomic system, the HS fcc solid, and 
the site-site radial distribution function for a molecular
system, the hard-dumbbells CP1 solid. We will
determine the structure both when one particle is fixed
and when all the particles are allowed to move. For the HS fcc solid, we considered
a simulation box with $N=$108 particles, so that the possible existence of
an inhomogeneity would result in an appreciable change in the radial
distribution function. As shown in Fig. \ref{fig_fdr} (a), the radial distribution
function is identical regardless of whether one particle is fixed or not.
As with regard to the hard-dumbbells CP1 solid, we considered a 
simulation box with only $N=$32 particles. 
In this case the center of mass of molecule 1 was fixed but 
molecule 1 was allowed to rotate. 
Our results show that the site-site
radial distribution function is again identical for a system
where all the particles are allowed to move and for a system where
the center of mass of one of the particles is fixed
(see Fig. \ref{fig_fdr} (b)). Note that it is important
that the dumbbell with fixed center of mass
is allowed to rotate. If molecule 1 is frozen at a given orientation,
the remaining molecules of the solid will not 'see' all
the possible orientations of molecule 1. 
Therefore, all the possible orientations
of the fixed molecule are not sampled and
the fixed particle will introduce an inhomogeneity
in the system. We checked that this is indeed true by computing
also the site-site radial distribution function for a system where one
particle is not allowed to translate and is not allowed to rotate. In this
case, it is observed that the value of the site-site radial distribution
function at contact is affected by the constraint on the orientation
of the carrier molecule. In particular, we obtained that the value
at contact is 5.072 when all the molecules are free to move,
which is equal (within statistical error) to the value at 
contact when the position of molecule 1 is fixed but it is
allowed to rotate, 5.070. However, when molecule 1 is not allowed to
translate nor to rotate, the contact value of the radial distribution
is somewhat lower (5.014), which means that the constraint on the 
orientation introduces
an inhomogeneity in the solid.

	The validity of the Einstein molecule approach for
molecular solids was checked by comparing the free energies of different
molecular solids with those obtained using the Frenkel-Ladd Einstein crystal
method (as implemented by Polson \emph{et al.}\cite{polson}).
At this stage, as the purpose was to show that both methods
lead to exactly the same results, we performed unusually 
long simulations in order to reduce the statistical error.
In what follows, we describe the simulation details for each model.

For the HD CP1 solid, we calculated the free energy
for a system with $N=$144 ($6\times 6\times 4$ unit cells) at
a number density $\rho^*=\rho\sigma_{HS}^3=0.590$, where $\sigma_{HS}$ is the
diameter of each one of the hard-spheres of a hard-dumbbell. 
As the solid is not cubic, we first
performed a Parrinello-Rahman\cite{parrinello-rahman,rao} NpT MC simulation consisting
of $5\times10^5$ MC cycles for equilibration plus another
$5\times10^5$ MC cycles for taking averages (a MC cycle is defined
as $N$ attempts to translate or rotate a particle plus one 
attempt to change the the matrix that defines the simulation
box). In agreement with previous calculations,
the Parrinello-Rahman NpT simulations show
that the ratio between the two edges of
the unit cell ($c/a$) is slightly different from that at close-packing.
Besides the changes in the shape
of the simulation box it is observed that the orientation
of the hard-dumbbells is also different from that at close-packing.
They change from $\theta =35.26^\circ$ to $\theta\approx 32^\circ$
and from $\phi=30^\circ$ to $\phi \approx 31^\circ $.
This has already been noted by 
Vega \emph{et al.}\cite{vega_jcp92} Once we have obtained the
equilibrium configuration at $\rho^*=0.590$, the free
energy was calculated by using 16 points to evaluate the
integral $\Delta A_2$ by the Gauss-Legendre quadrature formula. At each point,
the integrand of $\Delta A_2$ was evaluated by performing a NVT MC
simulation consisting 
of $2\times10^5$ MC cycles for equilibration and $2\times10^6$
MC cycles for taking averages 
at each value of the coupling parameter.
The term $\Delta A_1$ was calculated
in a simulation consisting of $2\times10^5$ MC cycles for equilibration and
$5\times10^5$ MC cycles for taking averages.
The maximum value of the coupling parameter used for the free
energy calculations was $\Lambda_E/(k_BT/\sigma_{HS}^2)=$4000.

For the patchy model we considered two system sizes $N=125$ and
$N=216$. In both cases, the
free energy was computed at the same thermodynamic state, 
$T^*=T/(\varepsilon_{LJ}/k_B)=0.2$ and 
$\rho^*=\rho \sigma_{LJ}^3 =0.763$, where $\varepsilon_{LJ} $ and $\sigma_{LJ}$
are the parameters of the LJ potential. The free
energy was evaluated by using 20 points to compute $\Delta A_2$,
and at each of those points we performed a simulation using $2\times10^5$
MC cycles for equilibration and $1\times10^6$ cycles for taking averages.
A maximum value of 
$\Lambda_E/(k_BT/\sigma_{LJ}^2)=\Lambda_{E,a}/(k_BT)=\Lambda_{E,b}/(k_BT)=$20000 
was used.

Finally, for ices XIII and XIV (two recently discovered solid phases
of water that exhibit both oxygen and proton ordering), we computed the free energy
at $p=1$ bar and $T=80K$. The simulation box contained
$3\times3\times2$ unit cells (504 molecules) in the case of ice XIII
and $3\times3\times5$ unit cells (540 molecules) for ice XIV.
As mentioned before, neither of these solid phases has cubic symmetry.
Ice XIII has a monoclinic unit cell and ice XIV has an orthorhombic
unit cell. Therefore, previously to the computation of the free energy,
the solid structure was relaxed to the equilibrium. For ice XIII,
we obtained that, at $p=1$bar and $T=80K$, the equilibrium simulation
box corresponds to $a=20.39$ {\AA}, $b=22.09$ {\AA} and $c=28.15$ {\AA}, 
and $\beta=109,6^\circ$ ($\alpha$ and $\gamma$ are equal to 90$^\circ$).
As ice XIV has
orthorhombic symmetry, only the length of the edges of the box were allowed
to fluctuate in the simulations, while the angles were kept fixed.
In this case, it was obtained that, at this thermodynamic state, the length of the
edges of the simulation box at equilibrium 
are $a=24.45$ {\AA}, $b=25.17$ {\AA} and $c=19.72$ {\AA}.
Once that the equilibrium shape of the simulation box was obtained,
the positions and orientations of the molecules (i.e, the positions of
the oxygens and hydrogens) in the reference
structure were taken from the crystallographic data provided 
by Salzmann \emph{et al.}\cite{iceXIII}
The NpT simulations consisted of $5\times 10^4$ MC cycles for equilibration
and $1.5\times 10^5$ cycles for taking averages. The free energy was computed
using a maximum value of 
$\Lambda_{E}/(k_BT/{\AA}^2)=\Lambda_{E,a}/(k_BT)=\Lambda_{E,b}/(k_BT)=$25000.
16 points were used to evaluate the term $\Delta A_2$ 
and, at each of these points, a simulation consisting of 
$5\times10^5$ MC cycles ($3\times10^4$ cycles for equilibration) was carried out, 
while the term $\Delta A_1$
was obtained from a simulation consisting of $1\times10^6$ MC cycles
($2\times10^5$ for equilibration).

	The results of the free energies for these systems, as calculated
using the Einstein crystal method and the Einstein molecule approach are
shown in Tables \ref{tbl_test} and \ref{tbl_ice_test}. In addition to the
total free energy, the value of the different terms that contribute to the
free energy in both methods are also shown.
It can be seen that both methods give the same value of the free energy
within the statistical uncertainty.
Although the contribution of the different terms that contribute
to the free energy is not the same when the center of mass is fixed or 
when molecule 1 is fixed, their sum is invariant.
It is interesting to note that the difference between $\Delta A_{2}^*/Nk_BT$
and $\Delta A_2/Nk_BT$ is about $\frac{3}{2N}lnN$ (see discussion in
Ref. \onlinecite{review}).
Therefore, our results show that, indeed, the Einstein molecule approach 
is a valid route to compute the free energy of molecular solids.

\subsection{Free energy of ices XIII and XIV.}

	As this is the first time that the free energies of ices XIII
and XIV are given, we decided to perform more extensive 
calculations in this case. The effect of the shape of the simulation box
on the free energy was also studied. Moreover, as mentioned before, it is
not obvious what orientation of the water molecules should be
chosen in the reference structure. For that reason, we decided to explore some
of the possible orientations to see whether this choice affects
the results of the free energy calculations.
The results presented in this section have been obtained from shorter
simulations than those of the previous section. Typically each simulation
consisted of $2\times 10^5$ MC cycles ($4\times 10^4$ for equilibration).
This is usually enough to obtain a reasonable accuracy.

	First we calculated the free energy of both phases using the
Einstein molecule approach in three
different thermodynamic states, namely, p=1bar and T=80K, p=1bar 
and T=250K and p=5000bar and T=80K, so that there are two points
along one isobar and two points along one isotherm. 
The values of the free energies at those thermodynamic states
are shown in Table \ref{free_energy_ices}.
These data will serve us to check that our calculations are thermodynamically 
consistent, i.e., that the results of the free energy 
calculations are the same as those obtained by thermodynamic 
integration of the equation of state. 
For ice XIII, through free energy calculations, 
we obtained $A_1(80K, 1bar)= -77.51(4) Nk_BT$,
$A_2(80K, 5000bar)= -77.39(4) Nk_BT$ and 
$A_3(250K, 1bar)=-18.51(4) Nk_BT$. Starting from $A_1$
and performing thermodynamic integration along this isotherm
we obtained that $A_2(80K,5000bar)=-77.40(6) Nk_BT$. 
The integration along the isobar starting from $A_1$ yields
$A_3(250K, 1bar)=-18.46(6) Nk_BT$. Both estimations are
in agreement with the results obtained by means of
free energy calculations.
A good agreement was also obtained for ice XIV. In this case,
the free energy calculations provide $A_1(80K, 1bar)=-77.82(4) Nk_BT$,
$A_2(80K, 5000bar)=-77.73(4) Nk_BT$, and $A_3(250K, 1bar)=-18.45(4) Nk_BT$.
Starting from $A_1$ and integrating along the isotherm 80K, 
we obtained that $A_2(80K, 5000bar)=-77.74(6) Nk_BT$,
and integrating along the isobar 1bar, it is obtained
that $A_3(250K, 1bar)=-18.52(6) Nk_BT$, again in good agreement
with the results of free energy calculations. 

	Once we have confidence on the reliability of
our calculations, we studied the effect of the shape of 
the simulation box on the free energy. For that purpose,
for ice XIV,
the free energy was also computed for a simulation box that
has been deformed from the equilibrium shape at T=80K and p=1 bar.
The length of the edges at equilibrium $L_{x,0}$, $L_{y,0}$
and $L_{z,0}$, were deformed to $L_x'=L_{x,0}\times \alpha$,
$L_y'=L_{y,0}\times \alpha$, 
and $L_z'=L_{z,0}/\alpha^2$, so that the density remains
invariant under this change of the simulation box.
Our results show that the free energy increases when 
the shape of the simulation box does not correspond to that
at equilibrium.
In particular, when a deformation defined
by $\alpha=0.96$ is
applied, the free energy increases from its value at
equilibrium $A_{sol}/(Nk_BT)=-77.82(4)$ to $A_{sol}/(Nk_BT)=-77.00(4)$.
An increase is also found when the edges are scaled with $\alpha=1.04$,
for which it was found that $A_{sol}/(Nk_BT)=-77.10(4)$.
This is the expected result, as the equilibrium
structure corresponds to a minimum of free energy, and
any perturbation will result in an increase of the free energy.
These results evidence the importance
of obtaining the equilibrium structure previously to the
computation of the free energy. Otherwise, we would be overestimating
the value of the free energy.

	As mentioned before, the positions of the oxygens
and hydrogens (i.e., the position and orientation of the water molecules)
in the reference structure to be used for the Einstein field
were taken from the experimental values (for ices XIII and XIV both
the oxygen and hydrogens are ordered). 
However, the experimental equilibrium positions and orientations of the
molecules will not be 
exactly equal to those of the potential model used in the
simulations. Moreover, it is possible that one would want to study
some solid for which there are no experimental data available.
This does not pose a problem, because, in principle, the free energy should not depend
on the precise location of the external field provided that the
field reflects the structure of the system. However, we wanted
to check that this was indeed true and we computed the free energy
using another reference structure.
In particular, we now used a reference
structure in which the water molecules are oriented
as to minimise the potential energy. This structure was obtained
by simulated annealing. Starting from a configuration in which
the simulation box corresponds to the equilibrium (as obtained from the
NpT simulation) and where the water molecules have the same positions
and orientation as those found in experiments,
we performed a quenching from 80K to 1K, using 6 intermediate
temperatures, and keeping the shape of the simulation
box constant along the whole simulation. 
At each one of the temperatures, the system was allowed
to evolve during $2\times10^4$ MC cycles.
To avoid translations of the system as a whole, we fixed
the reference point (but not the orientation)
of molecule 1 in the annealing. The structure obtained from the annealing
should be close to the minimum in the potential energy
(the minimum energy structure would be obtained at 0K,
at 1K the structure is likely to be not yet at the minimum\cite{juan}).
Using the structures obtained from 
simulated annealing, we calculated again the free energies of
ices XIII and XIV 
(see Table \ref{free_energy_ices}). It can be seen that the 
free energy is independent of the reference structure. 
As expected, the terms $\Delta A_1$ and $\Delta A_2$ take different values
depending on which reference structure has been chosen. However, their
sum is independent of the reference structure. The
term $\Delta A_1$ is close to the lattice energy and, therefore, it
is obvious that its value will depend on the reference structure.
On the other hand, $\Delta A_2$ is the integral from the real 
solid to the Einstein molecule
with intermolecular interactions. In this case, the fact of changing
the reference structure means that the integral is performed 
from a new starting point, which results on a different value 
of the integral $\Delta A_2$.
However, our results show that the changes in $\Delta A_1$ and $\Delta A_2$
cancel out and, therefore, the same value of the free energy is obtained
regardless of which reference structure has been used. 
Finally we also evaluated the free energy for a reference structure
in which the atoms are located at their average positions and orientations at that
thermodynamic state (these were obtained by averaging over
500 snapshots and readjusting the positions of the hydrogens to obtain
the bond and angle distances of the TIP4P/2005 model for each molecule),
and again the same value of the free energy (within statistical 
error) is obtained.
Obviously, it
is desirable that the reference structure is close to the equilibrium
structure, otherwise larger values of the coupling parameter would be needed
and this will result in a higher statistical error in the evaluation
of $\Delta A_2$.

\begin{figure}[!ht]
\begin{center}
\includegraphics[width=120mm,angle=0]{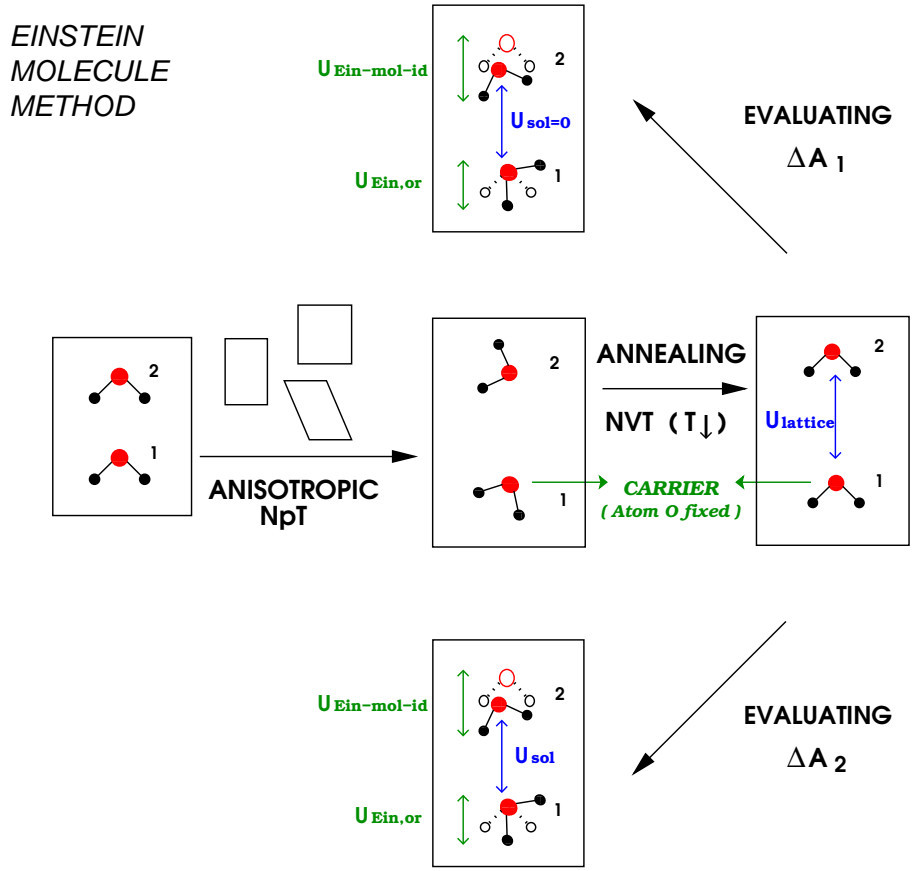}
\caption{\label{fig_esquema_emol} Schematic representation of the procedure
to compute the free energy of molecular solids by the Einstein molecule approach. 
In the first stage a Parrinello-Rahman
$NpT$ simulation is carried out to obtain the equilibrium shape of the
simulation box at the thermodynamic state under study. Second, starting
from a configuration with the equilibrium shape of the simulation box, the
position and orientations of the molecules in the lowest energy configuration
are obtained by simulated annealing (the shape of the simulation box
is kept constant during the quenching). In order to avoid translations of
the system as a whole, the position of the reference point of molecule 1
is kept fixed during the annealing. 
The final configuration obtained from this quenching, whose energy
is $U_{lattice}$, is then used
as the reference structure for the computation of the free energy
(i.e., for the evaluation of terms $\Delta A_1$ and $\Delta A_2$).
As described in the text, it is also possible to take the structure from
the experimental value of the coordinates of the molecules or
from the average positions.
To compute the term $\Delta A_1$, an NVT simulation of the ideal Einstein molecule
(i.e., with the position of the reference point of molecule 1 fixed)
is performed, along which the term $exp[-\beta(U_{sol}-U_{lattice})]$
is averaged, so that $\Delta A_1$ can be computed from Eq. \ref{eq_deltaa1}.
As with regard to the term $\Delta A_2$, 
several NVT simulations are performed where both the intermolecular potential 
and the Einstein field (for different values of the coupling
parameter $\Lambda'$) are present, in which again molecule 1 is not
allowed to translate. For each value of the coupling parameter,
the mean square displacement 
is averaged (i.e., for water it corresponds to the integrand of
Eq. \ref{eq_deltaa2_water}) and the integral 
Eq. \ref{eq_deltaa2_water} is evaluated to obtain the value
of $\Delta A_2$.
The term $\Delta A_2$ is then evaluated using Eq. \ref{eq_deltaa2}. }

\end{center}
\end{figure}

	Taking into account all these considerations, 
the procedure to compute the free energy
is schematically described in Fig. \ref{fig_esquema_emol}. 
It is important that, previously to the computation 
of the free energy, an appropriate
reference structure is obtained.

\begin{figure}[!ht]
\begin{center}
\includegraphics[width=85mm,angle=0]{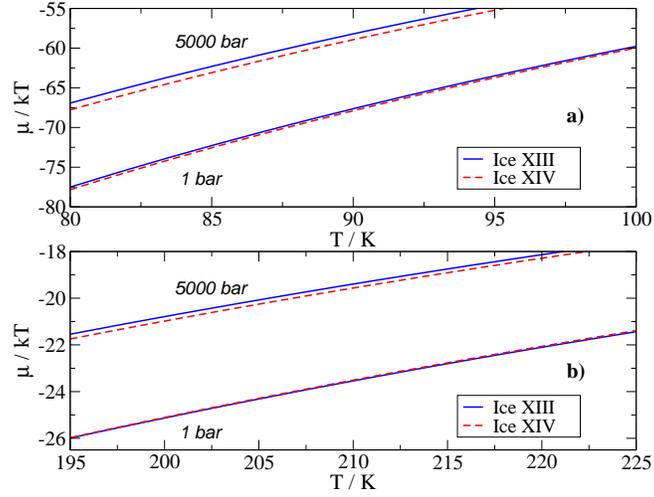}
\caption{\label{fig_coex_ice} (a) Chemical potential versus temperature for
ices XIII and XIV along the isobars p=1bar and p=5000bar at low
temperatures. (b) The same as (a) but at temperatures 
close to the melting point of the model.}
\end{center}
\end{figure}

	Before leaving this Section and because this is the first time 
that the free energy of ices XIII and XIV has been computed, we would
like to briefly discuss the relative stability of these two ice 
polymorphs. For that purpose, we computed the chemical potential
[$\beta \mu = (\beta A/N) + (P/\rho k_BT)$]
along the isobars p=1bar and p=5000bar by thermodynamic integration
(see Fig. \ref{fig_coex_ice}). It can be seen that, at $p=$5000bar ice XIV is
more stable than ice XIII at all temperatures,
i.e., from low temperatures up to melting
temperatures. On the contrary,
at p=1bar, ice XIV is slightly more stable than ice XIII
at low temperatures, but at temperatures close to melting 
ice XIII seems to be slightly more stable than ice XIV.
The phase transition seems to occur around  $T\approx 187K$.
In any case, it is important to note that
ice $I_h$ is the most stable phase at $p=1$ bar
for the TIP4P/2005 model.\cite{abascal05b}

\subsection{Size dependence of the free energy of molecular solids}

	Finally we also studied the size dependence of the free energy
of molecular solids by analysing the size behaviour 
of three different solid structures (sc, bcc, and fcc) 
for the octahedral patchy model. The free energies of 
those solid structures as obtained in this work using
the Einstein molecule approach for several values of $N$
are given in Table \ref{tbl_patchy}. Results for
the LJ fcc solid at $T^*=0.2$ and $\rho^*=1.28$ are also
given in Table \ref{tbl_lj}. In these calculations, the LJ potential
was truncated at a cutoff distance of 2.7$\sigma$ and 
long range corrections were used (obtained assuming $g(r)=1$ beyond the
cutoff). 
Our results show that for all the studied solid structures
the free energy exhibits a strong size dependence,
as found in previous 
studies.\cite{frenkel84,polson,barroso,enrique,vega_noya} 
It is interesting to note that 
the slope of the plot $A(N)/Nk_BT$ versus $1/N$ is different
depending on the solid structure, even for the same model potential. 
For the patchy model, we obtained that the slope is about
-14 for the sc structure, about -8 for the bcc and about -12 for
the fcc. This means that, in order to accurately calculate
the phase diagram of a given substance, a study of the system
size dependence must be performed for each considered solid structure,
which implies a large number of simulations.
Therefore it would be useful to have a simple
recipe to correct for the system size dependence as this could save
a large amount of computational time.

\begin{figure}[!ht]
\begin{center}
\includegraphics[width=85mm,angle=0]{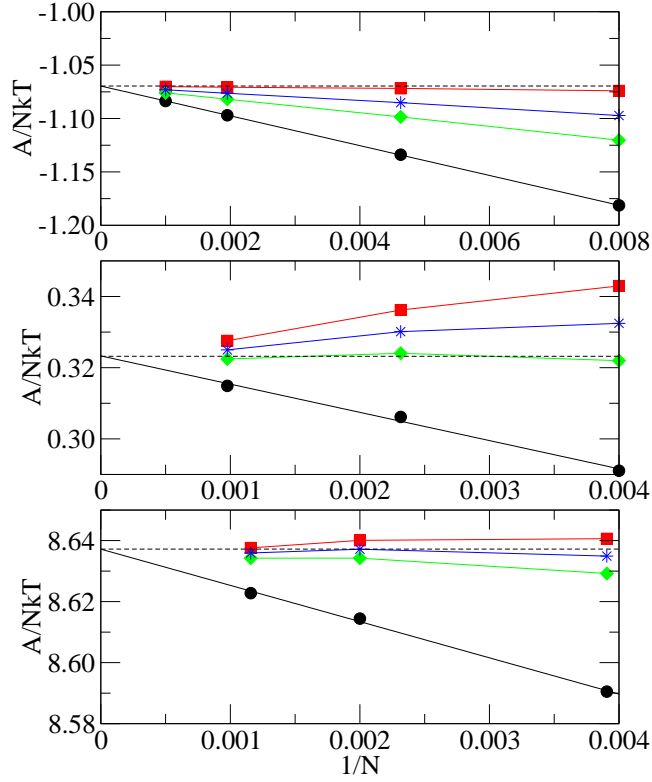}
\caption{\label{fig_patchy} (Colour online) Size dependence of the free energy 
of the sc (upper panel), bcc (middle panel) and fcc (botton panel)
solid structures of the six-patches octahedral model. Black circles
correspond to the true free energy of the system of size $N$
without any FSC correction and the black
line is a linear fit to these points. The black dashed-lines signals the
value of the free energy in the thermodynamic limit, obtained from the
fit. The red squares correspond
to the free energy corrected with the FSC-as1, the green diamonds
are the free energy corrected with the FSC-as2 and the blue stars
are the values corrected with the FSC-as3. The red, green and blue lines
are only a guide to the eyes.}
\end{center}
\end{figure}

	The performance of the FSC was studied for all the
considered solid structures of the octahedral anisotropic model.
In Tables \ref{tbl_patchy} and \ref{tbl_lj}, all the contributions to the
free energy are explicitly given, as this will allow us
to identify the terms that exhibit a stronger dependence with
the system size. 
The free energy obtained by applying the proposed FSC to the
free energy at a given $N$ for all the considered solid structures
of the octahedral patchy model are given in Table \ref{tbl_fsc} and
in Figure \ref{fig_patchy}. Results
of applying the FSC to the calculated free energies of the fcc LJ
solid are also given in Table \ref{tbl_fsc}.
It can be seen
that all the proposed recipes for finite size
corrections give a value of the free energy
closer to the thermodynamic limit than the estimate obtained
from the value of the free energy for a certain $N$. 
The FSC-HS, which was based on the slope of the
free energy as a function of $1/N$ for HS, obviously works
better when the slope is similar to the slope of HS
(around -7). The same is true for the FSC-HS2.
Therefore, at a given size, the prediction of the value of the
free energy in the thermodynamic limit for the sc and fcc structures
for the patchy model,
whose slopes were -14 and -12, respectively,
is not very accurate. The performance of the FSC-FL is also not
satisfactory. Although it also seems to give quite good
results in some cases (e.g., for the bcc solid in the patchy model), there are other
solids for which they correct only partially for the system
size dependence. Finally, the FSC-Asymptotic in its three variants
seem to give quite accurate results for all the cases studied.
This can be understood by looking at the size dependence of the
terms that contribute to the free energy (see Tables \ref{tbl_patchy}
and \ref{tbl_lj}). It can be observed that the terms $\Delta A_1$ and
the orientational contribution to $\Delta A_2$ ($\Delta A_{2,or}$) are 
almost independent of the system size. All the size dependence comes
from the free energy of the reference system (as given by Eq. \ref{eq_free_energy})
and, to a much lesser extent, from the translational contribution
to the integral $\Delta A_2$ ($\Delta A_{2,t}$). Therefore, by simply
taking the limit when $N \to \infty$ in this analytical expression
(Eq. \ref{eq_free_energy}),
the free energy at a given $N$ can be substantially corrected.

We also calculated the deviation from the correct value 
(i.e., the free energy
in the thermodynamic limit) for all the proposed FSC (see Table \ref{tbl_fsc_all}).
Data for the HS fcc solid at three different
thermodynamic states taken from a previous work\cite{vega_noya} are also included
in Table \ref{tbl_fsc_all}. The deviation from the correct value 
($d=\frac{A_{sol}(N)}{Nk_BT} - \frac{A_{sol}(N=\infty)}{Nk_BT}$)
was computed for the lowest system size studied in each case.
The mean deviation of each FSC computed as 
$\bar{d}= (\sum_{i=1}^n|d|/n)\times 1000$
is also given.
It can be seen that both the FSC-as1 and the FSC-as3 exhibit
the best performance, obtaining a mean deviation from the correct
value of 7 or 8 (in $10^{-3} Nk_BT$ units). The deviation of the rest of the FSC is not as
good, but still the mean deviation is typically around 14, which
is substantially lower than the mean deviation obtained from the
true value of the free energy at small values of $N$ (around 55).

\section{Conclusions}\label{conclusion}

	In this paper, we have extended the Einstein molecule
approach to the computation of the free energy of molecular solids.
The method has been tested using a variety of model potentials,
which include hard-dumbbells, the TIP4P/2005 water model and
a simple anisotropic model consisting of a spherical repulsive core
with some attractive sites. Our results show that both
the Einstein crystal method of Frenkel and Ladd (as corrected
by Polson \emph{et al.}\cite{polson}) and the Einstein
molecule approach of Vega and Noya give the same results of the free energy
within statistical accuracy.

	Once the Einstein molecule approach was tested,
this method was used to compute the free energies of
ices XIII and XIV for first time. The free energy was computed
at three different thermodynamic states, which allowed us to
test our free energy calculations by performing 
thermodynamic consistency checks.
In addition, we have stressed the importance of using the 
equilibrium shape of the simulation box in the computation of the 
free energy. Our results show that any deformation from this equilibrium
structure invariably leads to an increase of the free energy. This is
the expected behaviour, as the equilibrium structure is that that
minimises the free energy. Any deformation introduces stress
in the system that leads to an increase of the free energy.
Therefore, for solids with non-cubic
symmetry, it is important to perform a Parrinello-Rahman NpT
simulation to obtain the equilibrium shape of the simulation
box previously to the computation of the free energy.

	Moreover, we studied the effect that the choice of the 
reference Einstein field has on the calculation of free energies. 
In complex solids, such as, for example, ices XIII and XIV,
there is not an obvious choice of how the water molecules
should be oriented in the reference structure, as both solids
exhibit a complex unit cell with a large number of water molecules
and in which not all the molecules exhibit the same orientation.
We have performed calculations of the free energy of both
solid phases using a reference structure where the positions
and orientations of the molecules were taken from experimental
data and using a reference structure that has been obtained
by simulated annealing, i.e., using a reference structure
that minimises the potential energy (for the equilibrium
shape of the simulation box). Our results show that,
even though the two choices lead to different values of $\Delta A_1$ and
$\Delta A_2$, the addition of both terms is independent of the
choice of the reference structure. This is the expected result,
because we are computing the free energy of the same solid, but using
a difference reference system (i.e., the position and orientation of
the field are slightly different). Obviously it is desirable
to use a reference structure that is close to the minimum,
otherwise larger values of the coupling parameter will be needed
and this will result in a larger error in the evaluation of 
$\Delta A_2$. This is an important result, because, in many cases,
one will be interested in real solids with complex structures and
the choice of a reference structure will be a subtle issue. 
However, our results show that it is not necessary to obtain the
structure that minimises the potential energy, as far as 
the reference structure is not too far from this minimum.

	Finally, we have also studied the size dependence of 
the free energy for a simple anisotropic model. Our results
show that all the studied solid phases, namely, sc, bcc, and fcc,
exhibit a strong size dependence, although the slope of the plot
of A versus 1/N is different for each solid phase. In a previous work
we also found that the free energy of the fcc HS solid depends
slightly on the density.\cite{vega_noya} This means that
there is a complex dependence of the free energy with the system
size, which depends not only on the model potential, but also on
the thermodynamic state and on the solid structure. This result
seems to suggest that it might be difficult to obtain a simple recipe
that would allow us to obtain the free energy in the thermodynamic
limit from the calculated value at a finite size $N$. In any case, we
tested all the previously proposed FSC and we found that the
asymptotic FSC-as1 and FSC-as3 manage to give quite accurate 
estimates of the free energy in the thermodynamic limit 
for all
the solids studied so far. 

\acknowledgements
This work was funded by grants FIS2007-66079-C02-01 of Direcci\'on
General de Investigaci\'on and  S-0505/ESP/0229 from the Comunidad
Aut\'onoma de Madrid.
E.G.N. wishes to thank
the Ministerio de Educaci\'on y Ciencia
and the Universidad Complutense de Madrid for a Juan de la Cierva fellowship.
M. M. Conde would like to thank Universidad Complutense 
by the award of a PhD grant.

\newpage

\appendix

{\large\bf Appendix}
\vspace*{0.5cm}

	We will show that the free energy of the ideal Einstein
molecule (Eq. \ref{eq_free_energy}) can be obtained as a particular case of the ideal 
Einstein crystal with fixed center of mass.
In the ideal Einstein molecule the free energy is given by:
\begin{equation}
A_{sol} = A_{Ein-mol-id} +   \Delta A_1 + \Delta A_2 =  
	A_0 + \Delta A_1 + \Delta A_2
\end{equation}
The precise expression for $A_{0}$ is just that of $A_{Ein-mol-id}$
(see Eq. \ref{eq_free_energy}).
 
\begin{figure}[!ht]
\begin{center}
\includegraphics[width=85mm,angle=0]{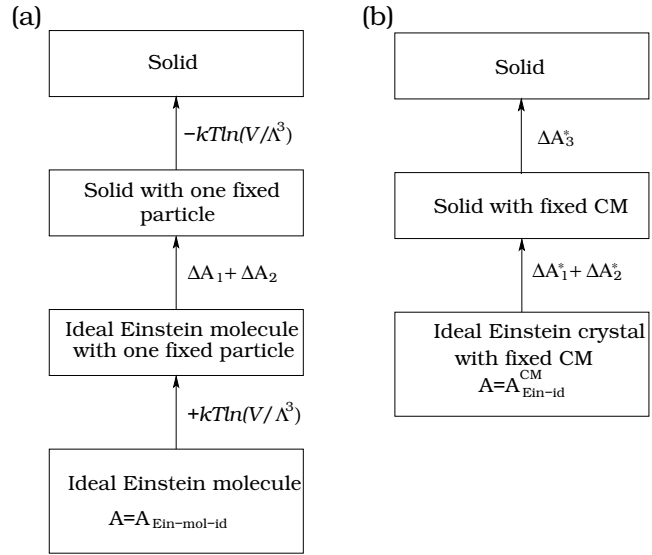}
\caption{\label{fig_esquema} Thermodynamic path used in (a) the Einstein
molecule approach\cite{vega_noya,review} and (b) the 
Einstein crystal method.\cite{frenkel84,polson}}
\end{center}
\end{figure}

In the Einstein crystal method the free energy is computed following the
integration path shown in Fig. \ref{fig_esquema} , so that the free energy
can be computed as:
\begin{eqnarray} 
A_{sol} & = & (A_{Ein-id}^{CM} + \Delta A_3^*) +   \Delta A_1^* + \Delta A_2^*  
	= A_0^* + \Delta A_1^* + \Delta A_2^* \\
          A_{0}^{*} &  = & A_{Ein-id}^{CM} +  \Delta A_3^*
\end{eqnarray}

In this appendix we will show that for a particular choice of the 
mass of the particles the Einstein crystal expression reduces to that
of the Einstein molecule expression.
 The term $A_{Ein-id}^{CM}$ is given by :
\begin{equation}
  A_{Ein-id}^{CM} = - k T ln ( Q^{CM}_{Ein,t} ) - k T ln ( Q_{Ein,or} )
\end{equation}
 where $Q^{CM}_{Ein,t}$ is given by (see Eq. 97 of Ref. \onlinecite{review}) :
\begin{equation}
\label{eq_q_eincris_cm}
Q^{CM}_{Ein,t}= P^{CM}(m_{1},...,m_{N}) \left(\frac{\pi}{\beta \Lambda_E}\right)^{3(N-1)/2}
\left(\sum^{N}_{i=1}\mu^{2}_{i}\right)^{-3/2},
\end{equation}
and  $P^{CM}(m_{1},...,m_{N})$ is the contribution of the momenta integral in
a system with fixed center of mass, where the dependence of 
$P^{CM}(m_{1},...,m_{N})$ on the masses is written explicitly. 
 The term $\Delta A_{3}^*$ is given by : 
\begin{equation}
\label{deltaa3}
\Delta A_{3}^*  = 
k_BT \left[ \ln ( P^{CM}(m_1,...,m_N)/P )  -   \ln (V/N) \right] 
\end{equation}
   where $P=1/(\prod\limits_{i=1}^N \Lambda_i^3)$ is the contribution to the space of momenta for
an unconstrained solid. 
 Putting together all terms contributing to $A_{0}^{*}$ one obtains:
\begin{equation}
A_{0}^{*}  = 
- k_B T ln \left(  \frac{1}{\prod\limits_{i=1}^N \Lambda_i^3}  \frac{V}{N} 
\left(\frac{\pi}{\beta \Lambda_E}\right)^{3(N-1)/2} 
\left(\sum^{N}_{i=1}\mu^{2}_{i}\right)^{-3/2} \right) - 
 k_BT ln ( Q_{Ein,or} )
\label{ao}
\end{equation}

Let us now compute $Q^{CM}_{Ein,t}$ for the case where all particles 
of the system have the same mass, $m_2=m_3= ... = m_N$, but
the mass of molecule 1 becomes infinitely large compared to that 
of the rest of the particles of the system (this is the choice of 
Vega and Noya\cite{vega_noya}). In this case $\mu_1=1$ and 
all $\mu_2=\mu_3=..=\mu_N=0$ (where $\mu_i=m_i/\sum_{i=1}^N m_i$). 
Obviously under these circumstances 
fixing the center of mass is equivalent to fixing the position of 
molecule 1 and, therefore, $\Delta A_1^{*} = \Delta A_1$ 
and  $\Delta A_2^{*} = \Delta A_2$. Let us see if for this 
particular choice we also obtain that $A_{0}^{*}=A_0$ which will complete the proof.
Substituting the reduced masses $\mu_1=1$ and
all $\mu_2=\mu_3=..=\mu_N=0$ in the expression of $A_0^*$, and
after some reordering of the terms, one obtains:
\begin{equation}
A_{0}^{*}  = 
k_BT \ln \left(\frac{N\Lambda^3}{V} \right) 
+ k_B T ln \left(\frac{\Lambda^2 \beta \Lambda_E}{\pi}\right)^{3(N-1)/2} 
- k_BT ln ( Q_{Ein,or} ) +k_BT \ln \left( \frac{\Lambda_1^3}{\Lambda^3} \right)
\end{equation}
which is exactly equal to the expression of $A_0$ in the Einstein molecule
method (Eq. \ref{eq_free_energy}), except for the trivial term 
$k_BT\ln \left( \frac{\Lambda_1^3}{\Lambda^3}\right)$, which
obviously will also appear in the fluid phase and will not
affect the phase equilibria. Thus we have proved that the Einstein molecule
method can be obtained as a limit case of the Einstein crystal method.


	We have seen that the precise value of $P^{CM}$ is irrelevant to 
compute the free energy (i.e., it does not appear in the expression of
$A^*$ as given by Eq. \ref{ao}). Nevertheless, for completeness, we will compute
its value for the two cases we are considering. We will start from the
general expression of 
 $P^{CM}(m_{1},...,m_{N})$:
\begin{equation}
\label{pcm}
P^{CM}(m_{1},...,m_{N}) =   \frac{1}{h^{3(N-1)}} \int \exp\left[-\beta \sum ^{N}_{i=1}
\frac{{\bf p}^{2}_{i}}{2m_{i}}\right]\delta(\sum^{N}_{i=1}
{\bf p}_{i})d{\bf p}_{1}...d{\bf p}_{N}
\end{equation}

 In the particular case that all particles of the system have the same mass
(this is the choice made by Polson et al.) then for all particles 
$m_{i}=m$ and $\mu_{i}=1/N$ and then:
\begin{equation}
P^{CM}(m,...,m) = \frac{1}{\Lambda^{3(N-1)}} N^{-3/2}
\end{equation}
and 
\begin{equation}
\label{eq_q_einmol_mol1}
Q^{CM}_{Ein,t}= \left( \frac{1}{\Lambda } \right)^{3(N-1)} 
\left(\frac{\pi}{\beta \Lambda_E}\right)^{3(N-1)/2} 
\end{equation}
See the derivation of this equation in Ref. \onlinecite{review} (Eq. 101).

Let us now compute $P^{CM}$ for the case where all particles 
of the system have the same mass, $m_2=m_3= ... = m_N$, but
the mass of molecule 1 becomes infinitely large compared to that 
of the rest of the particles of the system (this is the choice of 
Vega and Noya). In this case $\mu_1=1$ and 
all $\mu_2=\mu_3=..=\mu_N=0$. 

For this choice of masses:

\begin{equation}
\label{pcm2}
P^{CM}=   \frac{1}{h^{3(N-1)}} \int \exp\left[-\beta \sum ^{N}_{i=1}
\frac{{\bf p}^{2}_{i}}{2m_{i}}\right]\delta(
{\bf p}_{1})d{\bf p}_{1}...d{\bf p}_{N} 
\end{equation}
 where $\sum_{i=1}^N {\bf p}_i = 0$ was
simplified  to
${\bf p}_1=0$
when the mass of molecule 1 becomes infinitely large.
 
It is straightforward to integrate this expression to obtain:
\begin{equation}
\label{pcm3}
P^{CM}=   \frac{1}{h^{3(N-1)}} \left( \frac{2 m \pi}{\beta} \right)^{3(N-1)/2}
= \left( \frac{1}{\Lambda } \right)^{3(N-1)}
\end{equation}

	Therefore, the translational contribution to the partition
function is:
\begin{equation}
\label{eq_q_einmol_mol2}
Q^{CM}_{Ein,t}= \left( \frac{1}{\Lambda } \right)^{3(N-1)} 
\left(\frac{\pi}{\beta \Lambda_E}\right)^{3(N-1)/2},
\end{equation}
 which is identical to the expression obtained for the case where
all particles have the same mass (Eq. \ref{eq_q_einmol_mol1}). 
Therefore the expression for $A_{Ein-id}^{CM}$ is the same when 
all particles have the same mass or for the case where all have
the same mass but particle 1 which becomes infinitely heavy. 

	The partition function of an unconstrained Einstein crystal
is given by:
\begin{equation}
Q_{Ein} = \left( \frac{1}{\Lambda} \right)^{3N} 
\left( \frac{\pi}{\beta \Lambda_E} \right)^{3N/2}
\end{equation}
so that constraining the center of mass in the Einstein crystal
amounts to reducing the number of degrees of freedom by 3. 
Notice that this is not the same as $\Delta A_3^*$ as given
by Eq. \ref{deltaa3}. The reason is that Eq. \ref{deltaa3} gives the change in free energy
for fixing the center of mass in a system with translational invariance
(i.e., the energy of the system is invariant to a translation ${\bf \Delta}$
of all the particles), and such invariance has been used in the derivation
leading to the term $-\ln(V/N)$. Notice that the Einstein crystal does not
have translational invariance (the energy changes when all the
particles are translated by ${\bf \Delta}$ since the lattice does
not move), so that $\Delta A_3^*$ cannot
be used to get the free energy change for fixing the center of mass in
this case.

\newpage
\newpage


\newpage

\begin{table}[!h]
\small
\centering
\caption{\label{tbl_test} Free energy of the sc structure of the patchy
model particles  at $T^*=0.2$, and of the CP1 structure 
of hard dumbbells (HD), as obtained
using the Einstein molecule and the Einstein crystal methods.
For the patchy model we used $\Lambda_E/(k_BT/\sigma_{LJ}^2)=20000$ 
and $\Lambda=\sigma_{LJ}$ and
for HD $\Lambda_E/(k_BT/\sigma_{HS}^2)=4000$ and $\Lambda=\sigma_{HS}$.}
\begin{tabular}{lccccccccccc}
\hline\hline
      &  & & \multicolumn{4}{c}{Einstein molecule} &  & \multicolumn{4}{c}{Einstein crystal}\\
    \cline{4-7}  \cline{9-12}
System & $\rho^*$  &  $N$ & $\frac{A_0}{Nk_BT}$ & $\frac{\Delta A_1}{Nk_BT}$ 
&  $\frac{\Delta A_2}{Nk_BT}$ 
&  $\frac{A_{sol}}{Nk_BT}$ & & 
$\frac{A_0^*}{Nk_BT}$ & $\frac{\Delta A_1^*}{Nk_BT}$ &  $\frac{\Delta A_2^*}{Nk_BT}$
&  $\frac{A^*_{sol}}{Nk_BT}$ \\
\hline
Patchy (sc)   & 0.763 &   125   & 27.747 & -14.614 & -14.313 & -1.181(7) & &
27.689 & -14.614 & -14.256 & -1.181(7) \\
Patchy (sc)   & 0.763 &   216   & 27.792 & -14.614 & -14.311 & -1.134(7) & &
 27.755 & -14.614 & -14.278  & -1.138(7) \\
 \hline 
HD (CP1)      & 0.590  &   144  & 19.633 & 0.001 & -7.056 & 12.578(7) & & 
19.580 & 0.001 & -7.005 & 12.576(7)  \\
\hline \hline 
\end{tabular} 
\end{table}

\begin{table}[!hbt]\centering
\begin{center}
\caption{Free energies of ices XIII and XIV as calculated using the Einstein crystal and
the Einstein molecule methods. The simulation box contained 
$N=504$ water molecules for ice XIII and $N=540$ for ice XIV. 
Long simulations were performed
in order to reduce the statistical error. The maximum value of the coupling
parameter was $\frac{\Lambda_E}{k_BT}=$25000{\AA}$^{-2}$ and
we used $\Lambda=1$ {\AA}. The free energy was 
calculated by performing NVT simulations with the equilibrium simulation box 
at the studied thermodynamic state, namely, T=80K and p=1bar.}
\label{tbl_ice_test}
\small
\begin{tabular}{ccccccccccccc}
\hline
\hline
 & & &  & \multicolumn{4}{c}{Einstein molecule} &  & \multicolumn{4}{c}{Einstein crystal}\\
    \cline{5-8} \cline{10-13}
Ice   &
$p$(bar)  &
$T$(K) &
$\rho (g/cm^3)$ &
$\frac{A_{0}}{Nk_BT}$ &
$\frac{\Delta A_{1}}{Nk_BT}$ &
$\frac{\Delta A_{2}}{Nk_BT}$ &
$\frac{A_{sol}}{Nk_BT}$ & &
$\frac{A_{0}^*}{Nk_BT}$ &
$\frac{\Delta A_{1}^*}{Nk_BT}$ &
$\frac{\Delta A_{2}^*}{Nk_BT}$ &
$\frac{A_{sol}}{Nk_BT}$ \\
\hline
XIII & 1 & 80 & 1.262 &  29.491 & -91.229 &-15.769  &-77.508(8) & & 
29.472 & -91.229 & -15.756 & -77.512(8) \\
XIV & 1 & 80 & 1.332 &  29.493 & -91.073 & -16.259 & -77.839(8) & & 
29.475  & -91.073 &-16.246 & -77.843(8) \\
\hline
\hline
\end{tabular}
\end{center}
\end{table}
%
%
\begin{table}[!p]\centering
\begin{center}
\caption{Free energies of ices XIII and XIV as calculated using the
        Einstein molecule method. The data marked with an asterisk
        correspond to calculations of the free energy using a reference
        structure in which the positions and orientations of the Einstein
        field are those obtained from simulated annealing up to 1 K, while
        the data with two asterisks correspond to the structure with the average
        positions and orientations of the water molecules at the particular
        thermodynamic state.
As can be seen, the free energy does not depend on the choice
of the positions and the orientations of the Einstein external field.
In all these simulations we have taken $\Lambda=$1 {\AA}
and $\Lambda_E/(k_BT/\AA^2)=\Lambda_{E,a}/(k_BT)=
\Lambda_{E,b}/(k_BT)=$25000.}
\label{free_energy_ices}
\begin{tabular}{cccccccccc}
\hline
\hline
Ice & 
$p$(bar)  &
$T$(K) &
$\rho$(g/cm$^{3}$)  &
\large{$ \frac{U}{Nk_BT} $}  &
\large{$\frac{\Lambda_E}{k_BT}$}\normalsize{({\AA}$^{-2}$)} &
\large{$\frac{A_{0}}{Nk_BT}$} &
\large{$\frac{\Delta A_{1}}{Nk_BT}$} &
\large{$\frac{\Delta A_{2}}{Nk_BT}$} &
\large{$\frac{A_{sol}}{Nk_BT}$}\\
\hline
XIII & 1 & 80 & 1.262 & -89.08   &25000 & 29.49 & -91.23 & -15.77 & -77.51(4) \\
XIII$^*$ &1 & 80 &1.262  &-89.08 &25000 & 29.49 & -92.07 & -14.94 & -77.52(4) \\
XIII &5000 & 80 & 1.294 & -89.12 &25000 & 29.49 & -91.20 & -15.68 & -77.39(4) \\
XIII & 1 & 250 &1.208  &-26.01   &25000 & 29.49 & -28.96 & -19.04 & -18.51(4) \\
\hline
XIV & 1 & 80 & 1.332 &-89.64        & 25000 & 29.49 & -91.07 & -16.24 & -77.82(4) \\
XIV$^*$ &1 & 80 & 1.332 & -89.64    & 25000 & 29.49 & -92.61 & -14.72 & -77.84(4) \\
XIV$^{**}$ &1 & 80 & 1.332 & -89.64 & 25000 & 29.49 & -92.63 & -14.69 & -77.83(4) \\
XIV &5000 & 80 & 1.360 &-89.71      & 25000 & 29.49 & -91.02 & -16.20 & -77.73(4) \\
XIV & 1 & 250 &1.271  & -26.17      & 25000 & 29.49 & -28.95 & -18.99 & -18.45(4) \\
\hline
\hline
\end{tabular}
\end{center}
\end{table}

\newpage


\begin{table}[!p]
\centering
\caption{Free energies of the patchy model (see Eq. \ref{eq_patchy}) for different
values of $N$ and solid structures 
at $T^*=0.2$. 
We also report the value of the three different terms that contribute to
$\frac{A_{0,t}}{Nk_BT}$ (see Eq. \ref{eq_free_energy}), namely,
$\frac{A_{0,t,1}}{Nk_BT}=ln(\Lambda^3 \rho)/N$,
$\frac{A_{0,t,2}}{Nk_BT}=\frac{3}{2}ln(\Lambda^2 \beta \Lambda_E /\pi )$, 
$\frac{A_{0,t,3}}{Nk_BT}=-\frac{3}{2N}ln(\Lambda^2 \beta \Lambda_E /\pi )$.
In these calculations we used $\Lambda_E/(k_BT/\sigma_{LJ}^2)=20000$ 
and $\Lambda $ was taken as $\sigma_{LJ}$.}
\label{tbl_patchy}
\footnotesize
\begin{tabular}{cccccccccccc}
\hline\hline
System &  $\rho^*$  &  $N$  & $\frac{A_{0,t,1}}{Nk_BT}$ &
$\frac{A_{0,t,2}}{Nk_BT}$ & $\frac{A_{0,t,3}}{Nk_BT}$ 
& $\frac{A_{0,or}}{Nk_BT}$ & 
$\frac{A_{0}}{Nk_BT}$ & $\frac{\Delta A_1}{Nk_BT}$ & 
$\frac{\Delta A_{2,t}}{Nk_BT}$ & $\frac{\Delta A_{2,or}}{Nk_BT}$ & $\frac{A_{sol}}{Nk_BT}$ \\
\hline
Patchy (sc) & 0.763  &  125 & -0.002 & 13.138 & -0.105 & 14.716 & 27.746 & -14.614 & -5.731 & -8.583 & -1.181 \\
Patchy (sc) & 0.763  & 216 & -0.001  & 13.138 & -0.061 & 14.716 & 27.792 & -14.614 & -5.728 & -8.583 & -1.134 \\
Patchy (sc) & 0.763  & 512 & -0.001  & 13.138 & -0.026 & 14.716 & 27.828 & -14.614 & -5.729 & -8.582 & -1.097 \\
Patchy (sc) & 0.763  & 1000 & -0.000 & 13.138 & -0.013 & 14.716 & 27.840 & -14.614 & -5.729 & -8.581 & -1.084 \\
 \hline 
Patchy (bcc) & 1.175 & 250  & 0.001 & 13.138 & -0.053 & 14.716 & 27.802 & -13.718 & -5.231 & -8.562 & 0.291 \\
Patchy (bcc) & 1.175 & 432  & 0.000 & 13.138 & -0.030 & 14.716 & 27.824 & -13.718 & -5.236 & -8.564 & 0.306 \\
Patchy (bcc) & 1.175 & 1024 & 0.000 & 13.138 & -0.013 & 14.716 & 27.841 & -13.718 & -5.241 & -8.567 & 0.315 \\
\hline
Patchy (fcc) & 1.360 & 256 & 0.001 & 13.138 & -0.051 & 14.716 & 27.804 & -6.193 & -2.912 & -10.108 & 8.591 \\
Patchy (fcc) & 1.360 & 500 & 0.001 & 13.138 & -0.026 & 14.716 & 27.828 & -6.192 & -2.913 & -10.109 & 8.614 \\
Patchy (fcc) & 1.360 & 864 & 0.000 & 13.138 & -0.015 & 14.716 & 27.839 & -6.190 & -2.915 & -10.111 & 8.623 \\
 \hline\hline
\end{tabular}
\end{table}

\clearpage

%
\begin{table}[!h]
\caption{Value of the different terms that contribute to the free energy of the LJ
fcc solid
at $\rho^*=$1.28 and $T^*=$2.0. The LJ potential was truncated at
2.7$\sigma_{LJ}$. Long range corrections (assuming that g(r)=1 beyond the
cutoff) have been added.
We also report the value of the three different terms that contribute to
$\frac{A_{0,t}}{Nk_BT}$, namely,
$\frac{A_{0,t,1}}{Nk_BT}=ln(\Lambda^3 \rho)/N$,
$\frac{A_{0,t,2}}{Nk_BT}=\frac{3}{2}ln(\Lambda^2 \beta \Lambda_E/\pi)$, 
$\frac{A_{0,t,3}}{Nk_BT}=-\frac{3}{2N}ln(\Lambda^2 \beta \Lambda_E/\pi)$.
The free energy calculations
were performed using a maximum value of the coupling parameter 
$\Lambda_E/(k_BT/\sigma_{LJ}^2)=14000$. 
$\Lambda $ was taken as $\sigma_{LJ}$.}
\label{tbl_lj}
\footnotesize
\centering
\begin{tabular}{cccccccc}
\hline\hline
$N$ & $\frac{A_{0,t,1}}{Nk_BT}$  & $\frac{A_{0,t,2}}{Nk_BT}$ & 
$\frac{A_{0,t,3}}{Nk_BT}$ & $\frac{A_{0}}{Nk_BT}$ & 
$\frac{\Delta A_1}{Nk_BT}$ & $\frac{\Delta A_2}{Nk_BT}$ &  $\frac{A_{sol}}{Nk_BT}$ \\
\hline
256   & 0.001 & 12.603 & -0.049 & 12.555 & -3.620 & -6.365 & 2.570 \\
500   & 0.000 & 12.603 & -0.025 & 12.578 & -3.620 & -6.372 & 2.586 \\
864   & 0.000 & 12.603 & -0.015 & 12.589 & -3.620 & -6.377 & 2.592 \\
1372  & 0.000 & 12.603 & -0.009 & 12.594 & -3.620 & -6.380 & 2.594 \\
 \hline\hline
\end{tabular}
\end{table}

\newpage


\begin{table}[!h]
\caption{Free energies of the sc, bcc, and oriented fcc crystals for
the octahedral patchy model at $T^*=0.2$ and for the LJ model at $T^*=2.0$
including finite size corrections (FSC). No FSC corrections
means the true free energy for the system of size $N$.}
\label{tbl_fsc}
\footnotesize
\centering
\begin{tabular}{cccccccccccccccc}
\hline\hline
     &  &  & \multicolumn{9}{c}{$A/(NkT)$}  \\
    \cline{4-16} 
System & $\rho^*$ & $N$ & No FSC 
& &  FSC-HS2  && FSC-FL & & FSC-HS & & FSC-as1 && FSC-as2 && FSC-as3  \\
\hline
Patchy (sc) & 0.763 & 125 & -1.181 & & -1.123 & & -1.104 & & -1.125 & & -1.074 & & -1.120 & & -1.097 \\
Patchy (sc) & 0.763 & 216 & -1.134 & & -1.096 & & -1.084 & & -1.101 & & -1.071 & & -1.098 & & -1.085 \\
Patchy (sc) & 0.763 & 512 & -1.097 & & -1.079 & & -1.073 & & -1.083 & & -1.071 & & -1.082 & & -1.076 \\
Patchy (sc) & 0.763 & 1000 & -1.084 & & -1.073 & & -1.070 & & -1.077 & & -1.070 & & -1.076 & & -1.073 \\
Patchy (sc)  &  0.763 & $\infty$ &  {\bf -1.069} & & {\bf -1.069} & & {\bf-1.069} & & {\bf-1.069} & & {\bf -1.069} & & {\bf -1.069} & & {\bf -1.069} \\
 \hline
Patchy (bcc) & 1.175 & 250  & 0.291 & & 0.324 & & 0.335 & & 0.319 & & 0.343 & & 0.322 & & 0.332  \\
Patchy (bcc) & 1.175 & 432  & 0.306 & & 0.327 & & 0.334 & & 0.322 & & 0.336 & & 0.324 & & 0.330  \\
Patchy (bcc) & 1.175 & 1024 & 0.315 & & 0.325 & & 0.328 & & 0.322 & & 0.328 & & 0.322 & & 0.325  \\
Patchy (bcc) & 1.175 & $\infty$ &  {\bf 0.324} & & {\bf 0.324} &   &   {\bf 0.324}  &  &   {\bf 0.324} &  & {\bf 0.324} & & {\bf 0.324} & & {\bf 0.324} \\
\hline
Patchy (fcc) & 1.360 & 256 & 8.591 & & 8.623 & & 8.634 & & 8.618 & & 8.641 & & 8.629 & & 8.635 \\
Patchy (fcc) & 1.360 & 500 & 8.614 & & 8.633 & & 8.639 & & 8.628 & & 8.640 & & 8.634 & & 8.637 \\
Patchy (fcc) & 1.360 & 864 & 8.623 & & 8.634 & & 8.638 & & 8.631 & & 8.638 & & 8.634 & & 8.636  \\
Patchy (fcc) & 1.360 & $\infty$ &  {\bf 8.637} & & {\bf 8.637} & & {\bf 8.637} & & {\bf 8.637} & & {\bf 8.637} & & {\bf 8.637} & & {\bf 8.637} \\
\hline
LJ & 1.28 & 256  & 2.570 & & 2.602 & & 2.613 & & 2.597 & & 2.618 & & 2.593 & & 2.606 \\
LJ & 1.28 & 500  & 2.586 & & 2.605 & & 2.611 & & 2.600 & & 2.611 & & 2.598 & & 2.604 \\
LJ & 1.28 & 864  & 2.592 & & 2.604 & & 2.608 & & 2.600 & & 2.606 & & 2.599 & & 2.603 \\
LJ & 1.28 & 1372 & 2.594 & & 2.602 & & 2.605 & & 2.599 & & 2.603 & & 2.598 & & 2.601 \\
LJ & 1.28 & $\infty$ & {\bf 2.601} & & {\bf 2.601} & & {\bf 2.601} & & {\bf 2.601} & & 
{\bf 2.601} & & {\bf 2.601} & & {\bf 2.601} \\
 \hline\hline
\end{tabular}
\end{table}

\newpage

\begin{table}[!h]
\caption{Performance of the different FSC. Deviation of the corrected
values of the free energy at a given N from the value at the thermodynamic
limit
($d=\frac{A_{sol}(N)}{Nk_BT} - \frac{A(N=\infty)}{Nk_BT}$).
For the solids studied in this work we computed the mean deviation
for the smallest size studied, while for the HS solid it was estimated
for $N=$256.
The mean deviation for each FSC
is also provided (computed as $\bar{d}=\sum|d|/n\times10^3$).}
\label{tbl_fsc_all}
\small
\centering
\begin{tabular}{lccccccccccccccccc}
\hline\hline
System & $\rho^*$ & & No FSC & & FSC-HS2 & & FSC-FL & & FSC-HS & & FSC-A1 & & FSC-A2 & & FSC-A3 \\
\hline
HS & 1.04086  & & -0.028 & & 0.004 & & 0.016 & &  0.000 & & 0.003 & & -0.009 & & -0.003 \\
HS & 1.099975 & & -0.030 & & 0.002 & & 0.013 & & -0.003 & & 0.007 & & -0.008 & &  0.000 \\
HS & 1.1500 & & -0.034 & & -0.002 & & 0.009 & & -0.007 & & 0.002 & & -0.010 & & -0.004 \\
LJ & 1.2800 & & -0.031  & & 0.001 & &  0.012  & & -0.004 & & 0.017 & & -0.008 & & 0.005 \\
Patchy (sc) & 0.763 & & -0.112 & & -0.054 & & -0.035 & & -0.056 & & -0.005 & & -0.051 & & -0.028 \\
Patchy (bcc)& 1.175 & & -0.033 & &  0.000 & &  0.011 & & -0.005 & &  0.019 & & -0.002 & &  0.008 \\
Patchy (fcc)& 1.360 & & -0.046 & & -0.014 & & -0.003 & & -0.019 & &  0.004 & & -0.008 & & -0.002 \\
$\bar{d}$     &  & & 55  & & 11 & & 14 & & 13 & & 8 & & 14 & & 7 \\
 \hline\hline
\end{tabular}
\end{table}

\newpage

\clearpage


\newpage

\newpage

\newpage

\newpage

\end{document}